\pdfoutput=1

\documentclass[sigconf]{acmart}

\usepackage{amsmath,graphicx}
\usepackage{soul}
\usepackage{algorithm}
\usepackage{algorithmic}

\newcommand\wally[1]{#1}
\newcommand{\set}[1]{\mathbf{#1}}

\newcommand{\new}{\vspace{0.09in}}
\newcommand{\minus}{\vspace{-0.05in}}
\newcommand\mbf[1]{\mathbf{#1}}

\newcommand{\squishlist}{
	\begin{list}{$\bullet$}
		{  \setlength{\leftmargin}{0.8em}
	} }   
\newcommand{\squishend}{
\end{list}  }

\renewcommand\footnotetextcopyrightpermission[1]{} 
\setcopyright{none}

\acmDOI{}

\acmISBN{}

\acmPrice{}

\usepackage[normalem]{ulem}			
\usepackage[utf8]{inputenc}			
\usepackage{cleveref}
\usepackage{amsmath}
\usepackage{enumitem}
\usepackage{multirow, hhline}
\crefname{section}{§}{§§}
\Crefname{section}{§}{§§}

\graphicspath{{figs/}}

\makeatletter
\AtBeginDocument{\let\hl\@firstofone}
\makeatother

\usepackage[colorinlistoftodos]{todonotes} 
\usepackage{geometry}
\usepackage{marginnote}

 \usepackage{xcolor}

\begin{document}


\newcommand{\name}{{\em SubAoA}}


%

\title{\vspace{-0.00in} Multi-AoA Estimation under Real Multipath Channels \vspace{-0.3in}}
\title{\vspace{-0.00in} Estimating Multiple Angles of Arrival \\in a Steering Vector Space \vspace{-0.05in}}
\title{\vspace{-0.00in} Estimating Angles of Arrival (AoA) of Mutiple Echoes \\in a Steering Vector Space \vspace{-0.05in}}
\author{Yu-Lin Wei and Romit Roy Choudhury\\ }

\affiliation{
Department of Electrical and Computer Engineering, \\ 
University of Illinois at Urbana-Champaign}
\email{{yulinlw2,croy}@illinois.edu}
\maketitle

\section*{Abstract}

Consider a microphone array, such as those present in Amazon Echos, conference phones, or self-driving cars.
One of the goals of these arrays is to decode the angles in which acoustic signals arrive at them.
This paper considers the problem of estimating $K$ angle of arrivals (AoA), i.e., the direct path's AoA and the AoA of subsequent echoes. 
Significant progress has been made on this problem, however, solutions remain elusive when the source signal is unknown (such as human voice) and the channel is strongly correlated (such as in multipath settings).
Today's algorithms reliably estimate the direct-path-AoA, but the subsequent AoAs diverge in noisy real world conditions.
\new

We design {\name}, an algorithm that improves on the current body of work.
Our core idea models signals in a new AoA sub-space, and employs a cancellation approach that successively cancels each AoA to decode the next.
We explain the behavior and complexity of the algorithm from first principles, simulate the performance across a range of parameters, and present results from real-world experiments.
Comparison against multiple existing algorithms like GCC-PHAT, MUSIC, and VoLoc shows increasing gains for the latter AoAs, while our computation complexity allows real-time operation.
We believe progress in multi-AoA estimation is a fundamental building block to various acoustic and RF applications, including human or vehicle localization, multi-user separation, and even (blind) channel estimation.

\section{Introduction}
Angle of arrival (AoA) refers to the angle $\theta$ over which a signal arrives at a receiver.
In reality, a transmitted signal bounces off multiple surfaces and arrives at the receiver over multiple AoA angles $\{\theta_{1}, \theta_{2}, ... \theta_{N} \}$.
This paper aims to infer the first $K$ AoA angles, $\theta_{1:K}, ~ K<N $.
Such capabilities can be helpful to various applications.
For instance, a smart speaker like Amazon Echo may be able to infer the exact location of a user by reverse triangulating (or ray tracing) the AoAs of the voice \cite{dibiase2000high,VoLoc,262015}.
Self-driving cars may be able to detect sounds of oncoming vehicles even if those vehicles are around the corner, hence not visible to the camera or LIDAR \cite{autocar_acoustic1, uk-car-noise, bbc-car-noise}.
Multiple AoAs can also serve as useful information to multi-user detection, where a recording device is attempting to separate out voices in a meeting room \cite{kim2006blind}, or a RF spectrum-sensing radio is tracking multiple devices around it \cite{li2011bayesian,joshi2013pinpoint}.
This paper is aimed at solving an algorithmic question that could boost these and other applications.
We call our algorithm {\name}.

\new


While a rich body of past work (e.g., GCC-PHAT, MUSIC, JADE, ESPRIT, MVDR, and others \cite{GCC, MUSIC, tong1999joint, ESPRIT, benesty2000adaptive, JADE, capon1969high}) have explored this algorithmic problem space, most have either focussed on optimizing the line of sight (LoS) AoA $\theta_{1}$, or have made implicit assumptions such as impulse-like signals \cite{manocha_2018,showen2009acoustic}, multiple uncorrelated sources \cite{MUSIC}, co-prime channels \cite{tongBCISurvey98,tong1999joint}, etc.
This paper is an attempt to estimate $\theta_{1:K}$ in uncontrolled conditions, such as real-world multipath channels and with unknown acoustic source signals (such as human voice).
We are able to attain up to $K=4$  with an array of $M=6$ microphones.
\new

The key challenge in estimating $\theta_{1:K}$ can be viewed as a problem of estimating a specific type of delay from a mixture of delays.
To elaborate, consider a voice signal that arrives over a multipath channel to an array of microphones.
Three types of delays are in play:\hl{
(1) Voice signals exhibit strong auto-correlation, meaning that the signal changes slowly with time.
Thus, a delayed version of a signal $S$ looks similar to $S$.}
(2) Multipath causes copies of the same signal $S$ to add up with different delays.
(3) Finally, the signals arrive with relative delays across microphones, depending on the angle of arrival (AoA) of the signal.
The net received signal is thus a mixture of all these delays and an AoA estimation algorithm must isolate the $3^{rd}$ type of relative-delay from this mixture.
Observe that if the source signal is known~\cite{li2016norm}, or even if the source signal exhibits low auto-correlation (e.g., white noise)~\cite{tugnait2003channel}, the problem is solvable. 
With signals like human voice, none of these are true.
\new


Estimating $\theta_{1:K}$ has gained particular relevance in modern times, primarily due to advances in speech recognition and voice-based interfaces.
In past decades, significant work was accomplished around conference phones where the phone needed to recognize different voices and estimate their AoAs.
In such applications, since these voice signals across users were uncorrelated, the problem was easier.
On the other hand, acoustic communication modems, SONARs, ultrasound imaging, all developed sophisticated geometric signal processing techniques (including AoA), but had the flexibility to design the source signals.
Known source signals (such as pre-designed preambles) extended clear algorithmic advantages. 
Our problem of estimating $\theta_{1:K}$ for an unknown voice signal inherits the worst of both worlds.
Moreover, the solutions are expected to run in real-time (i.e., in the granularity of seconds), in noisy environments, and on small voice-enabled devices and cheap IoT radios.
{\name} needs to tackle these practical challenges. 
\new



\hl{The key idea we bring to the problem is that signals can be processed in a new AoA space  which is a space spanned by all the $360$ possible AoA vectors (that are known from basic geometry).
This is in contrast to the conventional signal space, which is sensitive to noise.}
We will explain this from ground-up, but our high level intuition is as follows.
We observe that received signals -- the direct path and the echoes -- become correlated when either their source data is correlated, or when their AoAs are from similar directions.
Our central contribution is that representing signals in an AoA sub-space can give ``immunity'' to correlations of the source data.
Said differently, while past techniques are crippled by two problems, we are crippled by one, at the cost of a slight increase in computation.
The net result is that, even though the multipath echoes are delayed copies of each other, it is still possible to decode their AoAs one by one.
Our algorithm is iterative in nature, where each AoA is decoded and then cancelled, so that the next AoA can be decoded from the residual (orthogonal) sub-space.
This iterative process continues until all the residual AoAs are comparable to noise, at which point no other AoA can be decoded. 
This determines the value of $K$.
\new 


In sum, the advantages of \name\ arise along $2$ dimensions:
(1) Compared to signal sub-space approaches like MUSIC (and its many variants) \cite{MUSIC,ESPRIT,xu2017weighted,180296}, the geometric AoA sub-space reduces the AoA's angular error for multiple echoes, particularly in noisy environments.
(2) In comparison to optimization based algorithms \cite{bciConvergence, JADE, VoLoc} that must solve non-convex minimization problems to {\em search} for multiple AoAs, \name\ benefits from combining the null-space and the AoA space to drastically reduce computational complexity.
This delivers the real-time requirement.
\new



{\em Finally, what have we lost in exchange for the gains?}
When AoA vectors are close to each other, i.e., $\big( \theta_i - \theta_j \big)$ is small, then only one AoA may get decoded (since the other AoA's projection to the null space is small).
However, at the expense of some computation complexity, it is possible to recover some of these AoAs (i.e., by explicitly searching for the last few AoAs).
{\name} exports this as a knob that system designers can configure based on their application's requirement.
\new

We evaluate {\name} through simulations and real-world experiments.
The real experiments are performed in various multipath settings (small apartments and large labs) using a \hl{6-microphone array} from SEEED \cite{ReSpeaker}, mounted on top of a Raspberry Pi.
We placed the device at different locations and gave voice commands, such as ``Hey Siri'', ``Ok, Google'', etc.
To derive ground truth (which is non-trivial for AoAs), we place a high quality ``reference'' microphone right next to the human's mouth, and use this recording as the source signal.
The known source signal permits channel estimation at each microphone, which yields the true AoAs.
We compare this ground truth against {\name} as well as a number of existing algorithms, including GCC-PHAT, MUSIC, and VoLoc.
For tests of robustness, and sensitivity to various parameters, we run MATLAB simulations that mimic our real experiment settings.
The main results from our experiments are as follows:
\new

\textbf{(1)} While all algorithms estimate the direct path AoA accurately, \name\ shows a marked improvement against others in estimating multipath AoAs. 
For instance, in a quiet lab, the $75^{th}$ percentile of $\Delta \theta_2$ with \name\ is  $11^{\circ}$, which is $13\%$ smaller than MUSIC and $56\%$ smaller than VoLoc.
The gain is greater in a low SNR regime: 
$\Delta \theta_2$ is reduced by $54\%$ and $76\%$ compared to MUSIC and VoLoc, respectively, yielding only $12^{\circ}$ error for the $75^{th}$ percentile in the presence of noise. 
\textbf{(2)} The algorithm completion time is $16$x faster with {\name} compared to VoLoc, and slightly slower than MUSIC.
In absolute numbers, {\name} completes in $0.54$ seconds on a $4$ quadcore Intel laptop, compared to VoLoc's $6.67$ seconds, and MUSIC's $0.21$ seconds.
\textbf{(3)} {\name}'s performance stays robust across different voice commands, users, SNRs, and background noises.
In sum, the contributions in this paper can be summarized as follows:

\begin{enumerate}
    \item We identify room for improvement in a classical area by treating signals in a new AoA sub-space, then iteratively canceling each AoA to decode the next.
    We show that up to $K=4$ AoAs can be reliably decoded, even in noisy, multipath environments (where echoes are strongly correlated).
    \item We show that the proposed ideas translate from algorithm to practice.
    Importantly, the accuracy and computation complexity inherit the good qualities of existing algorithms, although our {\name} algorithm is unlike either of them.
\end{enumerate}

This paper is focused mostly on the algorithm, its properties, and its performance in practical environments.
We will build up the foundations of AoA and beamforming from absolute first principles, with an aim to make the paper self-sufficient to researchers and practitioners, so they can use it in application domains like IoT, acoustics, wireless localization, and mmWave beamforming.
We are preparing to open-source our code and data for public use.






\section{Primer from First Principles}
\label{sec:primer}
This section explains the mathematical constructs (and intuitions) for {\name} and closely related papers (such as MUSIC, GCC-PHAT, MVDR, blind channel estimation, etc.).
Familiar readers can skip to Section \ref{sec:past}, or even to Section \ref{sec:algo}.
Also, from now on, we will denote the LoS AoA as $\theta_0$, first echo as $\theta_1$, and so on for symbol consistency.

%
\minus

\subsection*{Phase lag at nearby microphones}
\begin{figure}[h]
\minus 
    \includegraphics[width=0.39\textwidth]{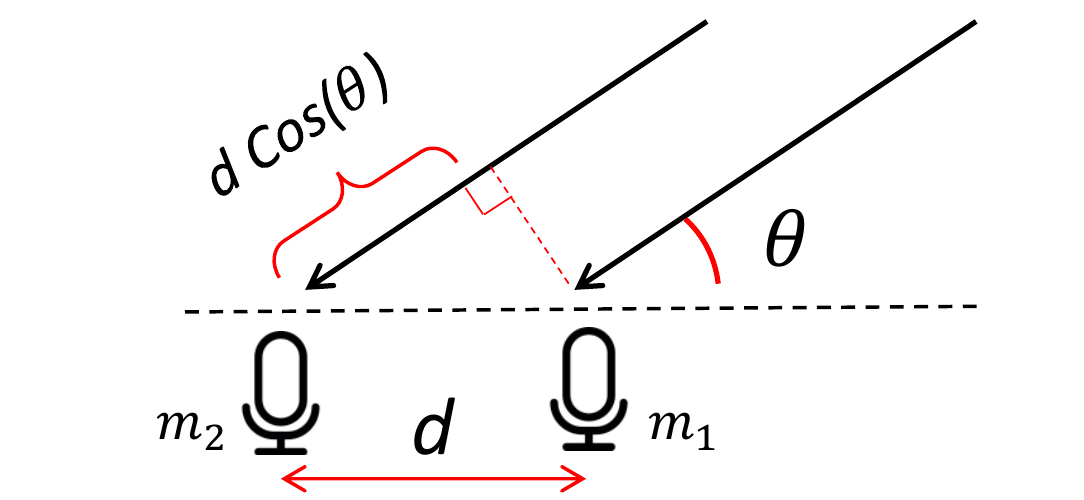}
    \minus 
    \caption{Angle of arrival $\theta$ causes signals to travel unequal distances at two separated microphones, causing a phase shift between the recorded signals as a function of $\theta$.}
    \vspace{-0.1in}
    \label{fig:toy}
    \minus
\end{figure}

Figure \ref{fig:toy} considers the toy case in which a signal $s$ from a transmitter arrives at a receiver composed of two adjacent microphones (separated by a distance $d$).
If the transmitter is far away, i.e., the transmitter-receiver distance $>>$ $d$, then the signal paths are almost parallel.
At any instant, the signals received by the two microphones are at a relative lag.
In this particular case, microphone $m_2$ receives the signal later than $m_1$ because the signal travels an additional distance to arrive at $m_2$.
This additional distance is $d Cos \theta$, where $\theta$ is the angle of arrival (AoA).
Translating this additional distance to phase, the phase difference $\Delta \phi$ between $m_1$ and $m_2$ is $\frac{2 \pi}{\lambda} d Cos \theta$.

\subsection*{AoA for Line of Sight}
Let us now generalize to $M$ microphones and assume there is no multipath (or echoes) from the environment.
If $m_1$ receives a signal $x_1$, we can say from above that the $r^{th}$ microphone $m_r$ receives a signal $x_r$ at a phase lag of $\Delta \phi_r = \frac{2 \pi}{\lambda} (r - 1) d Cos \theta  $ from $x_1$.
This means $\Delta \phi_r = (r - 1) \Delta \phi$.
In the frequency domain, this can be written as $X_r = X_1 e^{j (r-1) \Delta \phi}$, where $j$ is a complex number.
Writing this out as a vector for all the microphones, we have:

{
	\[
	\begin{bmatrix}
	X_1  \\ X_2 \\   \vdots  \\ X_{M}
	\end{bmatrix}
	=
	\begin{bmatrix}
	X_{1}  \\ X_{1} e^{j\Delta \phi}  \\   \vdots  \\  X_{1}e^{j (M-1) \Delta \phi}
	\end{bmatrix}
	= 
	\begin{bmatrix}
	1  \\ e^{j\Delta \phi}   \\ \vdots  \\  e^{j(M-1)\Delta \phi}
	\end{bmatrix}
	X_{1}
	\]
}

Given this vector $[X_1, X_2 ... X_M]^T$, decoding the AoA $\theta$ is easy.
The basic idea is to pretend the signal is arriving from some unknown AoA = $\psi$, subtract the corresponding delays from each microphone, and then add up the delayed signals as:
\minus

\begin{equation}
Sum_{\psi} ~=~ X_1 ~ + ~ X_2 e ^{-j\Delta \psi} ~+~ ... ~+~ X_M e ^{-j (M-1) \Delta \psi}
\label{eq:corr}
\end{equation}

From all values of $\psi \in [0, 2\pi]$, the one that results in the maximum $Sum$ gives us the correct $\phi$, which is then mapped to $\theta$.
The intuition is that when $\psi$ matches $\phi$, the relative delays between the microphone signals get compensated (or reversed).
Hence the signals becomes identical, meaning $X_1 = X_2 = X_3 ... = X_M$.
When these signals add up, they are called {\em coherent}, or {\em constructive}, or {\em in phase}, and their $Sum = M X_1$.
For incorrect $\psi$, the signals are not in phase and their sum is $< MX_1$.
\minus 

\subsection*{AoA under Multipath}
\begin{figure}[h]
\minus 
    \includegraphics[width=0.45\textwidth]{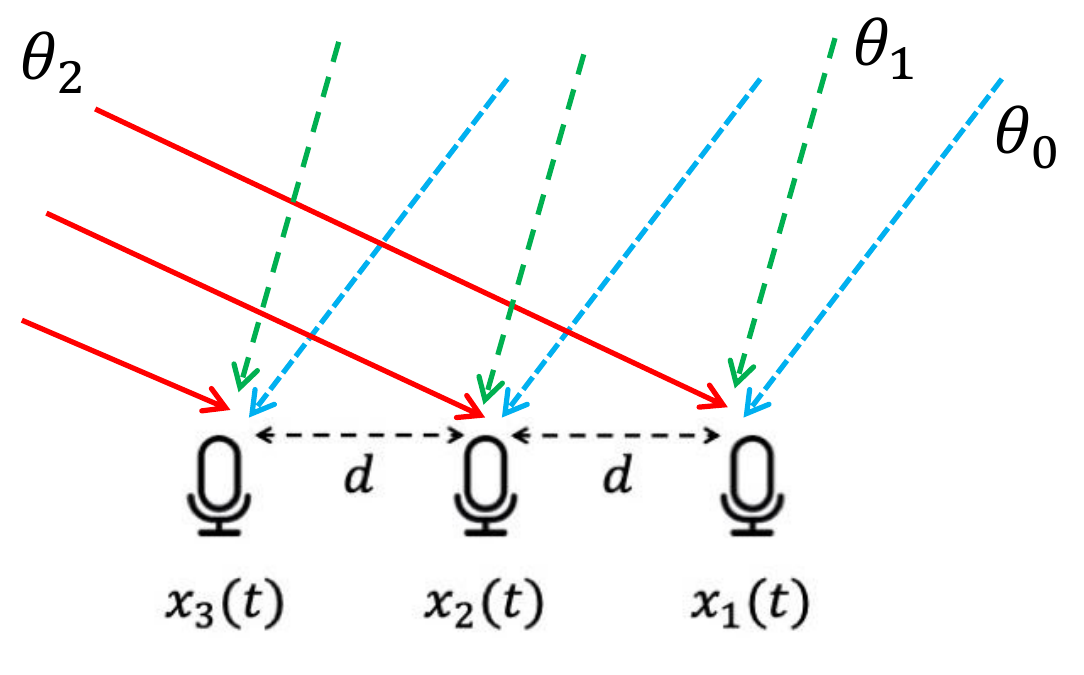}
    \minus  \minus \minus 
    \caption{LoS and multiple echoes add up at each microphone, each arriving from different AoAs.}
    \label{fig:toy-multi}
    \minus \minus 
\end{figure}

Figure \ref{fig:toy-multi} shows the multipath case where echoes of the source signal arrive at the receiver from different AoA angles.
Considering the first microphone $m_1$, the received signal is now a sum of multiple echoes, denoted as $X_1^{(e_0)}$ for the direct path, $X_1^{(e_1)}$ for the first echo, $X_1^{(e_2)}$ for the second echo, and so on.
The next microphone also receives each of these $K$ echoes but at corresponding delays, depending on that echo's AoA.
Thus, the received signal array (for $K$ echoes) will become:  
\minus \minus \minus 

{
	\[
	\begin{bmatrix}
	X_1  \\ X_2 \\  \vdots  \\ X_{M}
	\end{bmatrix}
	=
	\left[
	\begin{array}{cccc}
	1  			&     1  			&    \ldots  & 		1  \\
	e^{j\Delta \phi_{0}} 	&  e^{j\Delta \phi_{1}} 	& \ldots 	& e^{j\Delta \phi_{K}}      \\
	\vdots 			&   \vdots    			& 		& \vdots	\\
	e^{j(M-1)\Delta \phi_{0}} 	&  e^{j(M-1)\Delta \phi_{1}} 	& \ldots & e^{j(M-1)\Delta \phi_{K}}     \\
	\end{array}
	\right]
	\begin{bmatrix}
	X_{1}^{(e_0)}  \\  X_{1}^{(e_1)} \\ \vdots  \\  X_{1}^{(e_K)}
	\end{bmatrix}
	\]
}
\minus

Finally, the microphone hardware and the environment adds noise modeled as an additive noise vector $N$ on the right hand side of the above equation.
Algebraically, the equation can now be written as:
\minus \minus

\begin{equation}
X = AS + N
\label{eq:xasn}
\end{equation}

where $X$ is the ``received signal'' vector, $A$ is the ``steering. matrix'', $S$ is the ``source signal'' as measured at the reference microphone $m_1$, and $N$ is the ``noise'' vector.
The AoA question (in this multipath case) is aimed at estimating $\{ \Delta \phi_0, \Delta \phi_1,  ~ \ldots ~ \Delta \phi_K  \}$ first; knowing these, it is easy to compute $\{ \theta_0, \theta_1, ~ \ldots ~ \theta_K  \}$.

\subsection{Past AoA Algorithms} 
\label{sec:past}
 We sample $4$ key ideas from literature and the popular algorithms that emerged from them. 
\squishlist
\item[1.] Cross-correlation and GCC-PHAT \cite{GCC,benesty2000adaptive,brutti2011inference} 
\item[2.] Sub-space methods and MUSIC \cite{MUSIC,ESPRIT,xu2017weighted} 
\item[3.] Sparse recovery and BCI \cite{JADE,tong1999joint} 
\item[4.] Successive cancellation and VoLoc \cite{VoLoc,gollakota2008zigzag}
\squishend


\textbf{$\blacksquare$ Cross-correlation and GCC-PHAT} \\
Observe that Equation \ref{eq:corr} can be written as a dot product as follows:
\minus \minus \minus 

\begin{equation}
Sum_{\psi}
= 
 [1 \hspace{0.1in} e ^{-j\Delta \psi} \hspace{0.1in} \ldots \hspace{0.1in} e ^{-j (M-1) \Delta \psi}] ~ . 
 \begin{bmatrix}
	X_1  \\ X_2 \\  \vdots  \\ X_{M}
	\end{bmatrix}
\label{eq:corr2}
\end{equation}

This is a correlation between an AoA vector (of phase $\Delta \psi$) against the measured signal $X$ at the receiver.
A simple cross-correlation method (called Delay-and-Sum \cite{elko1996microphone}) attempts to perform such correlations for all values of $\Delta \psi \in [0, \pi]$ and the values of $\Delta \psi_i$ that show high correlation gives all the $K$ AoAs.
The intuition is that for a specific $\Delta \psi_i$, one of the echoes will add up coherently, and the result would be high even though other echoes are adding in-coherently.
As an analogy, if one is listening to an orchestra of many instruments, it might be possible to listen to the guitar by searching for the guitar pattern.
Correlating with $\Delta \psi$ is like listening to the guitar amid the other instruments.
\new

GCC-PHAT refines this intuition but generalizes it by removing the effect of magnitude.
In other words, GCC-PHAT does not want a louder violin to drown the guitar, so it normalizes the correlation function by the amplitudes of each instrument.
The hope is that each instrument can now be identified by the patterns they create in time (and not in loudness).
Mathematically, this normalization is performed in the denominator (the numertator is basic cross-correlation between $A$'s $i^{th}$ column and $X$, all in the frequency domain):
\minus 
\begin{equation}
Sum_{\psi} = \frac{A_i(\Delta \psi)^H X}{| A_i(\Delta \psi)^H X|}
\end{equation}

The AoAs correspond to those values of $\Delta \psi$ that produce peaks in $Sum_{\psi}$.
The issue is, violins, cellos, and several other instruments can often add up to sound like guitars, so this technique is often unreliable in real scenarios.
\new

\textbf{$\blacksquare$ Sub-space methods and MUSIC} \\
\begin{figure}[h]
\minus 
    \includegraphics[width=0.95\columnwidth]{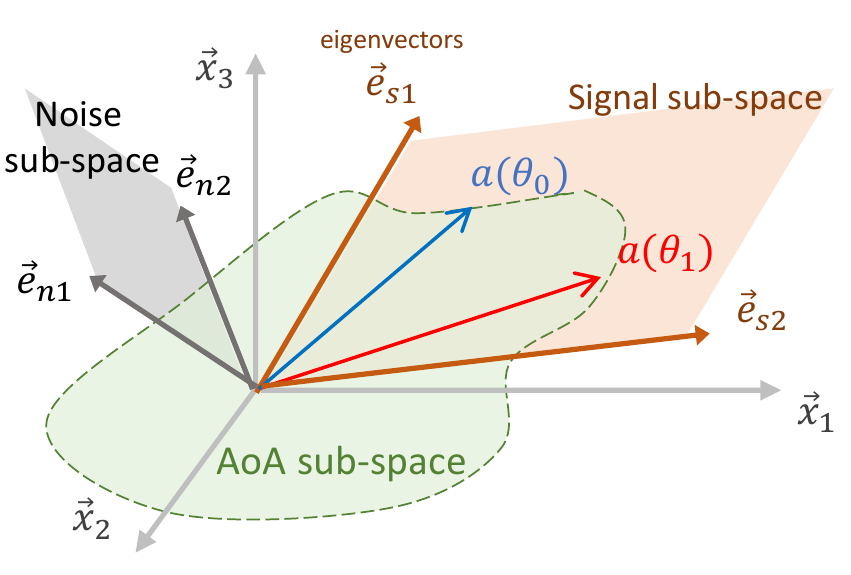}
    \minus  \minus \minus  \minus 
    \caption{Subspace method computes noise space by eigen-decomposition~\cite{MUSIC}.}
    \minus  \minus 
    \label{fig:MUSIC_idea}
    \minus
\end{figure}
These methods look at the AoA problem through the lens of linear algebra and vector spaces.
Their core intuition is that signals received on $M$ microphones form an $M$ dimensional space, however, assuming there are only $K<M$ echoes, the signals from all these echoes should span a $\mathbb{R}^K$ sub-space inside $\mathbb{R}^M$.
Said differently, the remaining $\mathbb{R}^{(M - K)}$ space should be spanned only by noise, and importantly, should be orthogonal to the signal sub-space.
Thus, if signals are represented in the appropriate sub-spaces, then orthogonality between signals and noise can reveal the AoA vectors.
\new

Mathematically, MUSIC realizes this intuition by computing the auto-correlation of $X = AS + N$, which leads to:
\minus 

\begin{equation}
R_{xx} = A R_{ss} A^H + \sigma^2 I
\end{equation}

Eigen-decomposition of the $R_{xx}$ matrix reveals the noise space, and it can be shown that different columns of $A$, i.e., the correct AoA vectors, are {\em orthogonal} to the eigenvectors of noise. 
~\wally{as shown in Figure ~\ref{fig:MUSIC_idea}.} 
Returning to our orchestra analogy, MUSIC shows that the cacophony (or noise) created by an amateur orchestra is a function of the instruments that are being played.
So, whether a guitar (or a cello) is part of that orchestra can be determined by analyzing the noise generated by the band.
\new

The issue is that MUSIC requires the musical instruments to be all different from each other.
Mathematically, this means that the $K$ source signals $[X_{1}^{(e_0)} ~~  X_{1}^{(e_1)} ~~ \ldots ~~  X_{1}^{(e_K)} ]$ must all be uncorrelated.
Unfortunately, in a multipath channel, each of these signal components is the echo, or delayed version of each other; hence they are highly correlated.
This derails MUSIC.
While ``spatial smoothing'' optimizations have attempted to alleviate the issue, they are effective only in limited scenarios.
\new 

\textbf{$\blacksquare$ Blind Inference and Sparse Recovery} \\
Blind channel inference (BCI) is a relatively modern idea that attempts to estimate the channel blindly, i.e., without knowing the source signal $s$.
The core idea essentially models the received signals at $2$ microphones as:
\minus
\begin{equation}
y_1 = h_1*s ~~~~~ \text{and} ~~~~~ y_2 = h_2*s ~~~ \implies Y_1 H_2 = Y_2 H_1
\end{equation}

\begin{table*}[t]
\begin{tabular}{|l|l|l|c|c|c|c|}
\hline
\multicolumn{1}{|c|}{\textbf{Algorithm}} & \multicolumn{1}{c|}{\textbf{Cite}} & \multicolumn{1}{c|}{\textbf{\begin{tabular}[c]{@{}c@{}}Target  Applications\end{tabular}}} & \textbf{\begin{tabular}[c]{@{}c@{}}Handles \\ correlated?\end{tabular}} & \textbf{\begin{tabular}[c]{@{}c@{}}K \\ AoAs?\end{tabular}} &  \multicolumn{1}{c|}{\textbf{Complexity}} & \textbf{\begin{tabular}[c]{@{}c@{}}Real-\\ Time?\end{tabular}} \\\hline
Delay and Sum  & \cite{elko1996microphone,antonio2011delay,varma2002time} 
& SONAR 
& Maybe & No & $O(M^2N)$ &Yes \\ \hline

GCC-PHAT  & \cite{GCC,benesty2000adaptive,brutti2011inference} 
& Voice Assistants 
& Maybe & No & $O(M^2N)$ & Yes \\ \hline

MUSIC & \cite{MUSIC,ESPRIT,xu2017weighted,180296} 
& Conference calls, Beamforming, WiFi
& No & Yes & $O(M^2N)$ & Yes \\ \hline

BCI and MLE & \cite{JADE,tong1999joint} 
& De-reverberation, Source separation 
& No & Yes & $\Omega(M^2N^2)$ & No  \\ \hline

VoLoc & \cite{VoLoc,gollakota2008zigzag} 
& Indoor Sensing, Voice localization 
& Yes  & Yes  & $\Omega(M^2N^2)$ & No  \\ \hline

SubAOA  & \bf{This work} 
& Voice localization, Beamforming
& Yes  & Yes  &$O(M^2N)$ & Yes \\ \hline
\end{tabular}
\caption{Comparison of existing AoA algorithms: $M$ denotes microphone array size and $N$ denotes signal length}
\minus \minus 
\label{tbl:alg}
\end{table*}

\begin{figure*}[h]
\minus 
    \includegraphics[width=0.3\textwidth]{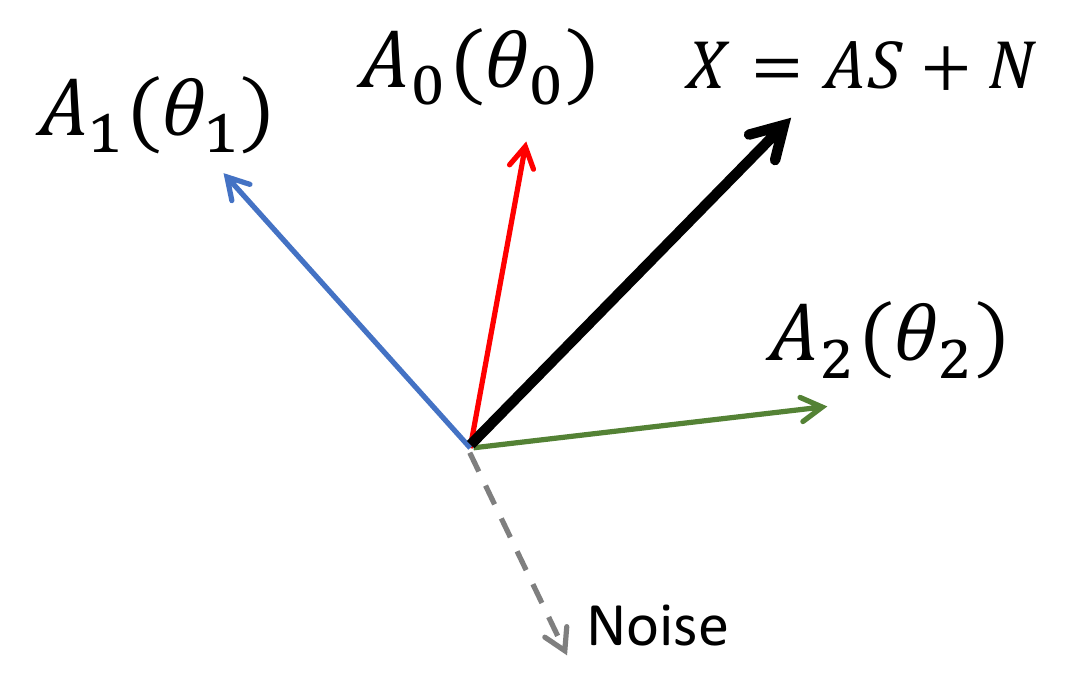}
    \includegraphics[width=0.3\textwidth]{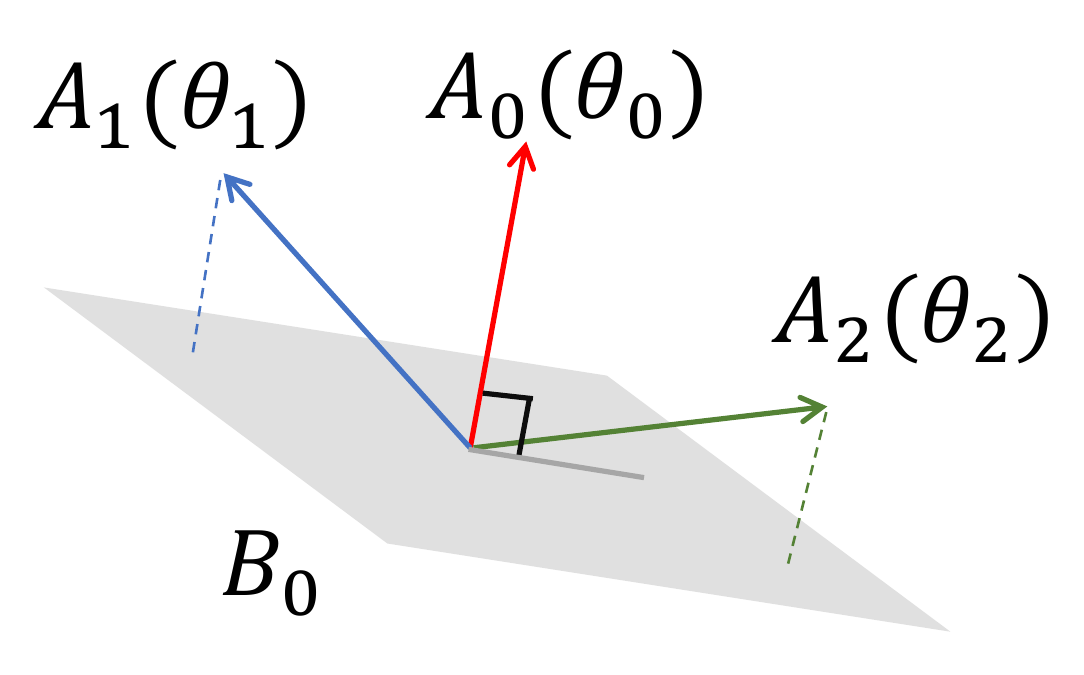}
    \includegraphics[width=0.3\textwidth]{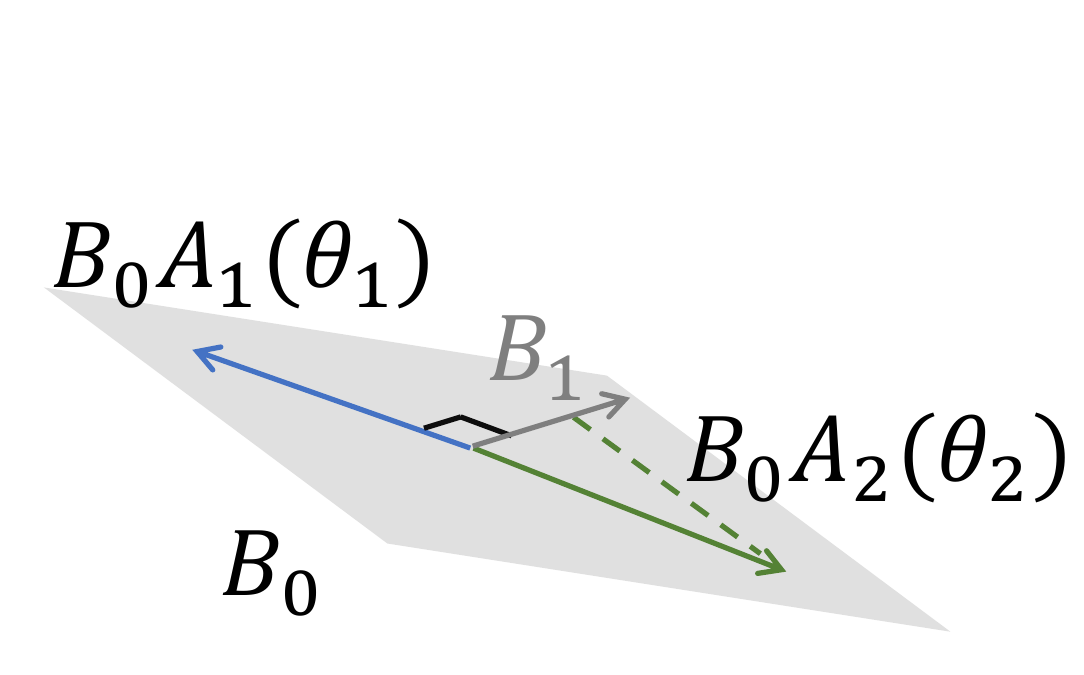}
    \minus 
    \caption{\wally{(a) X=AS + N expressed in the vector space. (b) Projection on the B0 sub-space orthogonal to $AoA_0$ (c) The projected signal does not contain $AoA_0$}}
    \minus \minus 
    \label{fig:B0-project}
    \minus
\end{figure*}

In other words, since the source signal $s$ is the same for both the microphone recordings, is it possible to solve for both the channels from the right side equation $Y_1H_2 - Y_2H_1 = 0$.
If $H_1$ and $H_2$ are decoded, then the AoAs can be estimated by computing the relative delays between the same channel taps.
Solving this is difficult in general, but assuming sparse channels (i.e., an $L_0$ norm optimization), it might be possible to recover $H1$ and $H_2$.
However, $L_0$ norms are intractable, so $L_1$ norms have been used.
Still, the computation cost is extremely high, and with hardware noise, the algorithms remain far from practice.
Table ~\ref{tbl:alg} \hl{presents a brief comparison between these techniques.}
\new

\textbf{$\blacksquare$ Iterative cancellation and VoLoc} \\
In 2019, VoLoc identified the opportunity that the LoS path is initially clean (i.e., unpolluted by echoes), so it might be possible to cancel the LoS after the first echo arrives.
This should give the first echo, which can then be cancelled out when the second echo arrives.
This iterative cancellation can reveal the echoes one by one, alongside their AoAs.
\new

The issue is that perfect cancellation is a heavy (non-convex) optimization since the decision variables are every possible delay, as well as amplitudes of the signal.
Moreover, with noise, the cancellation will never be perfect, indicating that the errors of cancellation will accumulate through each step.
Finally, the initial LoS path, although unpolluted, might be very weak for human voices (since humans cannot produce loud sounds suddenly).
This derails cancellation.
\minus

\section{Proposed Algorithm: {\name}}
\label{sec:algo}
Our central idea draws inspiration from MUSIC and VoLoc.
We begin with an intuitive explanation of {\name}, followed by a formal description.
\minus 

\subsection{Intuition and Visualization}
Figure \ref{fig:B0-project}(a) visualizes Equation \ref{eq:xasn}: $X = AS + N$.
Observe that $X$ is a vector formed by a weighted combination of \wally{$K$} AoA vectors, plus noise.
Now, using a sub-space approach like MUSIC, the LoS path AoA$_0$ (=$~\theta_0~$) can be decoded reliably from received signal vector $X$.
Knowing AoA$_0$, {\name} computes its orthogonal sub-space of dimensions $(M-1)$.
Let's denote this orthogonal sub-space as $B_0$ (Figure \ref{fig:B0-project}(b))


Observe that the product $B_0 X$ is a signal vector that has essentially cancelled out the LoS path (since they are orthogonal).
Viewed another way, $B_0 X$ is the residue signal after all AoA vectors (weighed by the echoes) have been projected to the $B_0$ space.
Similarly, $B_0 A$ is a sub-space in which the projected AoA vectors lie.
Thus, we have reduced the original problem to a smaller problem in a $(M-1)$ dimensional space, and importantly, this sub-problem does not contain the LoS signal component.
Our goal is to now decode the first echo, i.e., AoA$_1$ (=$~\theta_1~$), by applying the same method for AoA$_0$.
We then derive the orthogonal space $B_1$ for this echo, and then compute $B_1 B_0 X$ which eliminates both AoA$_0$ and AoA$_1$.
Iteratively, {\name} yields the AoAs of $K$ echoes until the AoA vector projections are comparable to noise, at which point no AoAs can be detected anymore.
\new


\textbf{Two behaviors of the algorithm are worth discussing:} \\
\textbf{(1) Transforming both Signal and AoA Space:} 
From Figure \ref{fig:B0-project}(c), the projected AoAs are in the $B_0$ sub-space, so when one of those AoAs is decoded, it should not be the same as the original AoA (before projection).
However, recall that {\name} also projects AoA-space with $B_0$ (in addition to $B_0 X$).
This transformation preserves the correct AoA vectors, which can thus be decoded to $\Delta \phi$ and ultimately $\theta$.
\new

\textbf{(2) Order of AoA Estimation:} 
If AoA$_1$ is angularly close to AoA$_0$, then AoA$_1$'s projection onto sub-space $B_0$ would be small.
However, the AoA$_2$ vector might exhibit a stronger projection on $B_0$.
Thus, in {\name}'s second step, AoA$_2$ will be decoded before AoA$_1$.
In the subsequent iteration (i.e., upon projection to AoA$_3$'s orthogonal space), AoA$_1$'s projection may prove stronger than AoA$_3$'s projection.
AoA$_1$ will get decoded then.
In general, the order of decoding AoAs in {\name} is a function of the power of that echo, as well as the angular separation with other echoes.
This is the reason why the acoustic channel cannot be immediately estimated from the AoAs alone.
As an analogy, channel estimation can be viewed as a global ordering of tap-delays across all microphones, while AoA only gives a partial ordering between microphone pairs.
However, knowing AoAs can help in blind channel estimation, a topic that we leave to future work.
\minus

\subsection{Algorithm Details}
Algorithm ~\ref{alg:subaoa} presents the pseudo code for \name.
We denote the received signal matrix as $\mbf{X}=[\mbf{x}_1,...\mbf{x}_M]$,  
where $\mbf{x}_m$ is the signals recorded by the $m^{th}$ microphone.
The steering matrix is denoted as $\mbf{A} = [\mbf{a}_1,...\mbf{a}_{360}] \in \mathbb{C}^{M\times 360}$, and $\mbf{a}_\theta$ is the steering vector of angle $\theta$. 
Note that the steering matrix $\mbf{A}$ varies across different frequencies, but without loss of generality, we explain for a single frequency.
Next, let's zoom into the details of the algorithm, containing essentially $2$ main steps. 
\new


{\bf Step 1: Compute noise space $N$ for max. likelihood AoA.} \\
The $M$-dimensional received signal is composed of the $K$-dimensional signal sub-space and the $(M-K)$-dimensional noise sub-space.
We apply principle component analysis (PCA) on the received signal matrix, and extract the $M-K$ least significant eigen-vectors to obtain the noise space $\mbf{N}$:
\minus \minus 

\begin{equation}
    \mbf{N}=PCA(\mbf{X})[:,-(M-K):end]
\minus 
\end{equation}
where $K$ is the expected number of signal paths. 
\new


Our next step is to compute the likelihood of each AoA angle by testing its orthogonality with $\mbf{N}$. 
A signal along a correct AoA angle will be strongly orthogonal to the noise space (close to zero).
The negative log-likelihood is defined as $p_\theta = -\log{\mathbf{a}_\theta^H\mathbf{N}\mathbf{N}^H\mathbf{a}_\theta}$. 
For steering vectors orthogonal to $\mbf{N}$, this negative log-likelihood will be maximized. 
We compute the noise space and AoA likelihood across all frequencies, sum up the likelihoods, and output the $\theta_0$ with the max likelihood. This $\theta_0$ is the LoS AoA estimation. 
\new 

{\bf Step 2: Compute the residual signals by removing signals arriving from the decoded AoA direction(s)} \\
Thus, once $\theta_0$ is known, we intend to cancel out from $\mbf{X}$, all the signal power arriving from this AoA $\theta_0$.
To this end, we project $\mbf{X}$ to $\mbf{B}_0\in\mathbb{C}^{(M-1)\times M}$, the null space of $\mbf{a}_{\theta_1}^H$.
\begin{align}
    \mbf{B}_0 &= N(\mbf{a}_{\theta_0}^H)^H\\
    \mbf{R} &= \mbf{X}\mathbf{B}_0^H
\end{align}
Naturally, the residual matrix $\mbf{R}$ does not contain any power from the LoS path after the projection. 
Then we also project the steering matrix $\mbf{A} = norm(\mbf{B_0}\mbf{A})$.
This gives both the received signal residual and the steering matrix residual, and importantly, both are free of $a_{\theta_0}$ components. 
We now plug these matrices back to {\bf Step 1}, replace $\mbf{X}$ with $\mbf{R}$ and estimate the next AoA $\theta_{1}$.
This iteration continues $K$ times till $\theta_K$. 


\begin{algorithm}[t]
\caption{SubAoA $(\mathbf{X}, \mathbf{A})$ }
\begin{algorithmic}
\STATE Initialize the residual $\mathbf{R}_0$ with the received signal $\mathbf{X}$ 
\FOR {the $k^{th}$ path $k=0:K-1$} 

\STATE Compute the orthogonal sub-space $\mathbf{N}$ by PCA on the residual $(\mathbf{R}_k)$
    \FOR {all possible AoAs $\theta$} 
        \STATE Find the steering vector $\mathbf{a}_{\theta}$ perpendicular to $\mathbf{N}$
    \ENDFOR
    \STATE $\theta_k = \arg\max_\theta -\log(\mathbf{a}_\theta^H\mathbf{N}\mathbf{N}^H\mathbf{a}_\theta)$
    \STATE Define the orthogonal residual sub-space $\mathbf{B}_k$ where $\mathbf{B}_k\bot\mathbf{a}_\theta$
    \STATE Project both the residual and steering matrix to $\mathbf{B}_k$:  $\mathbf{R}_{k+1}=\mbf{R}_k\mathbf{B}_k^H$, $\mathbf{A}=norm(\mathbf{B}_k\mathbf{A})$
\ENDFOR
\end{algorithmic}
\label{alg:subaoa}
\end{algorithm}

\subsection{Computational Complexity}
We analyze and compare the complexity of \name\ with MUSIC, GCC-PHAT, and VoLoc\footnote{ 
We omit the FFT complexity $O(N\log N)$ and assume all the received signals are already transformed to the frequency domain.}.
For each iteration, \name\ executes (a) PCA in $O(M^3+M^2N)$, (b) likelihood calculation in $O(M^2)$ and (c) residual projection in $O(M^2N)$, where $N$ is the number of received samples.
The total complexity of \name\ is $K\times\text{((a)+(b)+(c))}=O(KM^2N)$.
\new

For MUSIC, the complexity is only a single round of (a)+(b) = $O(M^2N)$.
For GCC-PHAT, the complexity of a microphone pair is $O(N)$, and with the GCC component repeated for all $M^2$ microphone pairs, the complexity is also $O(M^2N)$.
The complexity of VoLoc depends on the resolution of parameter search.
For each candidate location, the likelihood computation is $O(K^3N)$, since a matrix inversion is needed to cancel the paths. 
The total complexity is $O(R^2K^3N)$ where $R$ is the location resolution.
\new

In sum, \name\ features the same order complexity as MUSIC and GCC-PHAT ($K$ times higher, but the path count $K$ is small in real-world).
All of these algorithms are much more efficient than successive cancellation methods like VoLoc.

\subsection{Algorithm Analysis}

We discuss various properties, tradeoffs, and differences of \name\, especially in contrast to sub-space algorithms like MUSIC and their variants. 
\new



\textbf{Coping with correlated signals (multipath)} \\
In multipath environments, the signals for all $K$ paths are correlated (since they are delayed copies of the same source).
Therefore, for conventional sub-space methods like MUSIC, the signal sub-space will be less than $K$, and the noise space will be larger than the ideal rank, $M-K$. 
When we choose the least $M-K$ eigenvectors out from the noise sub-space, part of the signal sub-space gets removed (note that even with unsupervised clustering of eigenvalues, the problem is not resolved since weak signals will cause small eigenvalues).
Worse, a strong noise component will be selected into the signal space. 
In sum, decoding AoAs of correlated signals with MUSIC is similar to estimating AoAs under a noisy environment.
In such cases, weak echoes may not be orthogonal to the noise space, so MUSIC (and all its variants) will suffer for AoAs with weak echoes. 
\new

To mitigate this, \name\ estimates {\em only the strongest path} in each iteration. 
Now, consider an echo that is correlated to the LoS signal from the first iteration.
Since \name\ projects the AoA, a correlated echo will not be affected so long as it is angularly separated from the LoS AoA.
In other words, the residual $R_1$ does not contain any signals from the $\theta_0$ angle, i.e., the $\mbf{a}_{\theta_0}$ axis is removed. 
When we compute the new nosie space using this residual, the correlated LoS signal is absent, so the reflected path will dominate the residual, resulting in a large likelihood for $\theta_1$. 
\new

\begin{figure}[ht]
\minus \minus 
    \includegraphics[width=0.49\columnwidth]{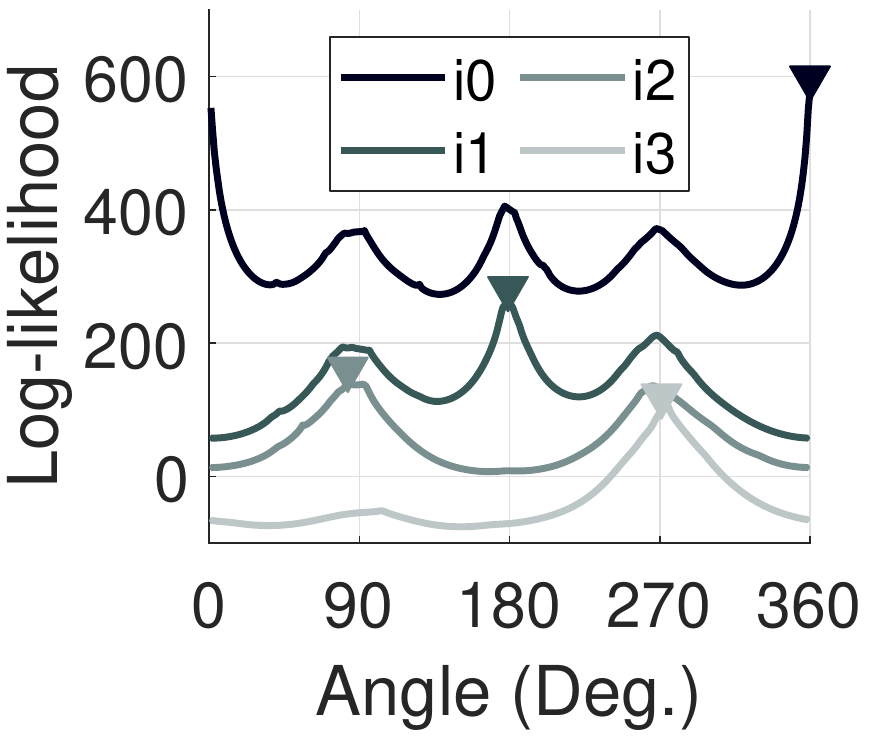}
    \hfill
    \includegraphics[width=0.49\columnwidth]{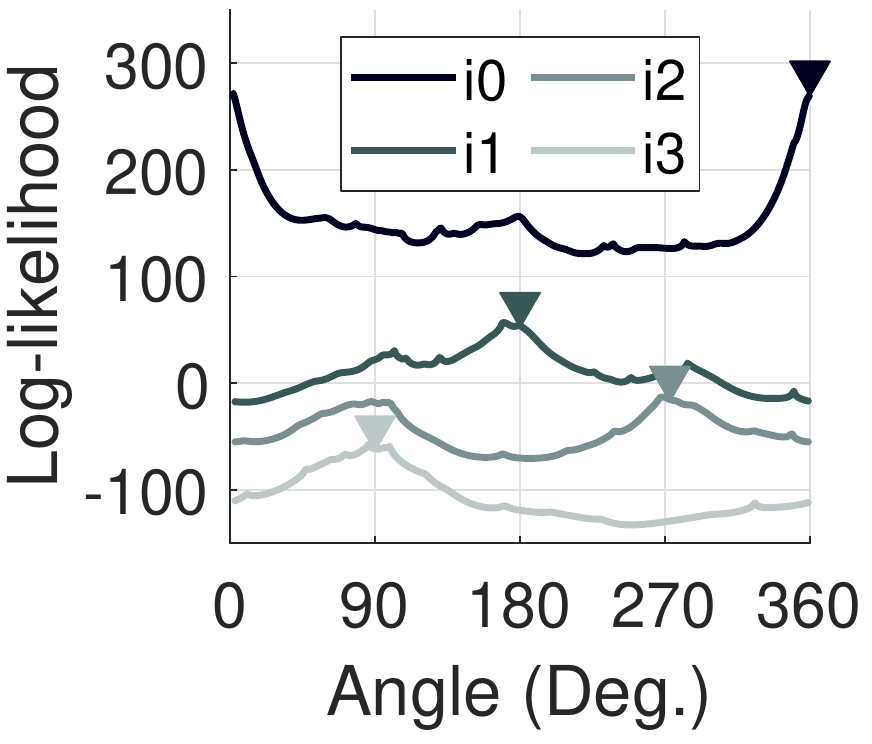}
    \minus \minus 
    \caption{AoA Likelihood where (a) white noise as the signal in the noise-free environment, and (b) speech as the signal in a noisy (SNR=10dB) environment.}
    \label{fig:simu_spec}
    \minus \minus \minus 
\end{figure}

Figure ~\ref{fig:simu_spec} shows MATLAB simulations where each line in the graph corresponds to one iteration of {\name}.
The solid triangle shows the AoA detected in that iteration, since that is the maximum likelihood peak.
Since {\name}'s first iteration is the same as MUSIC, the line $i0$ is actually MUSIC's performance. 
Observe that in noise-free environments (Figure ~\ref{fig:simu_spec}(a)), MUSIC and \name\ are comparable.
The difference becomes pronounced with noise; at SNR=$10dB$, Figure ~\ref{fig:simu_spec}(b) shows how \name\ preserves sharp AoA peaks while MUSIC falters.
This is evidence that AoA vectors are preserved in {\name}'s residual space (after projection),  allowing them to be decodable one by one.
We will show similar results from real-world testbeds as well.
\new



\new
\textbf{From signal sub-space to AoA sub-space} \\
A large class of sub-space algorithms (inspired by MUSIC) \cite{MUSIC,ESPRIT,xu2017weighted} computes the noise space from the signal covariance matrix. 
This requires the source signals to be uncorrelated to guarantee a $K$-ranked signal sub-space. 
In contrast, \name\ performs the eigen-decomposition on the steering matrix space $\mbf{A}$, rather than the signal space $\mbf{X}$.
This relaxes the assumption of uncorrelated signals; this highlights the core contribution of operating in the AoA sub-space. 
\new

\begin{figure}[ht]
    \includegraphics[width=0.9\columnwidth]{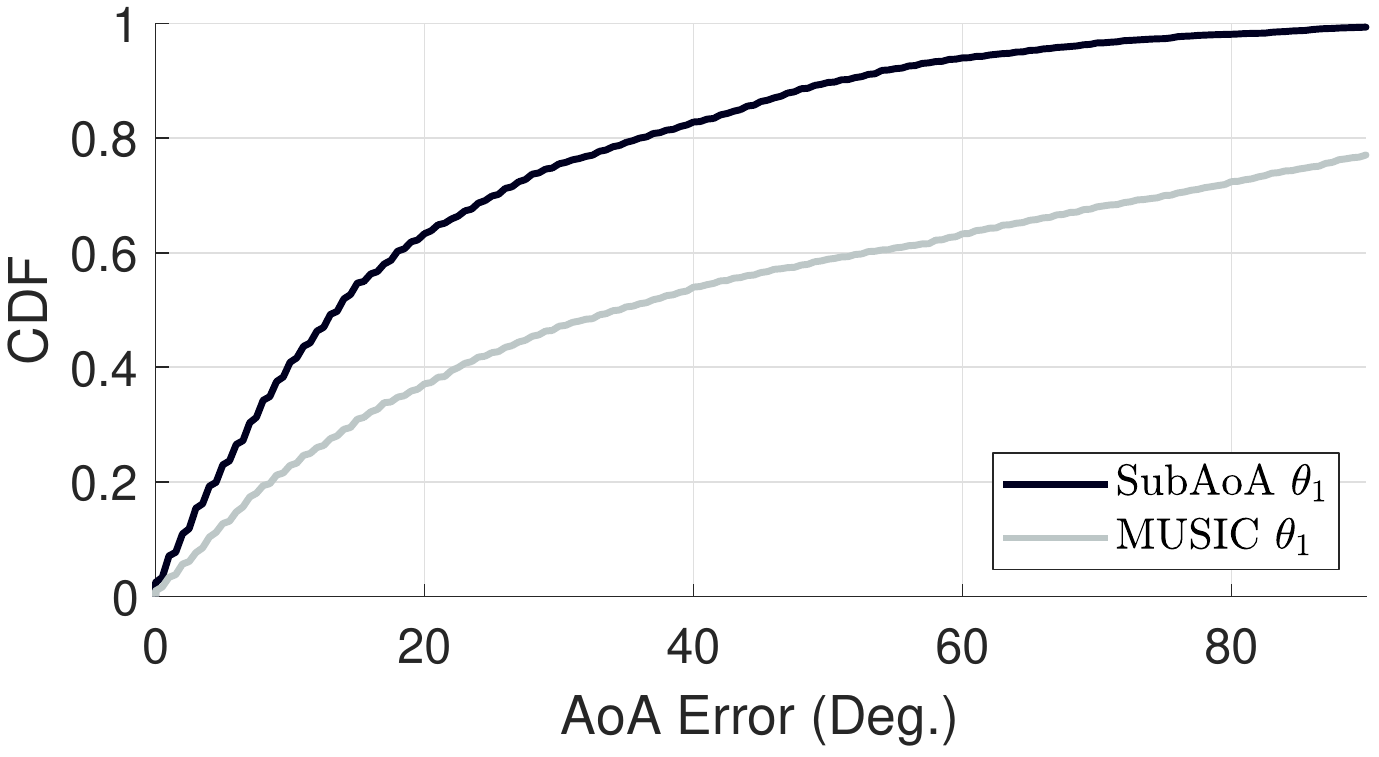}
    \minus
    \caption{Error compared with the MUSIC algorithm}
    \label{fig:simu_music}
    \minus \minus \minus 
\end{figure}

Figure~\ref{fig:simu_music} visualizes the gain of AoA sub-space across more than $100$ simulated multipath settings. 
We use real recorded speech as the source signal and set SNR = $10dB$; a room impulse response (RIR) outputs the indoor channel.
\name\ consistently outperforms MUSIC -- {\name}'s median $\theta_1$ error is $58\%$ lower because the echoes are strongly correlated.
Real testbed results will further corroborate these outcomes.
\new


\textbf{Upper bounds on $K$: in theory and practice} \\
In each iteration of {\name}, the rank of the residual reduces by $1$ due to projection.
In noise-free conditions, \name\ can estimate up to $M-1$ AoAs. 
However, the estimation error grows high after $\theta_4$ both in simulation and real-world scenarios (discussed later). 
The key reason is ``error accumulation'', i.e., AoA estimation error from previous iterations affects the residual in latter rounds. 
Specifically, if LoS AoA $\theta_0$ is estimated with a $5^\circ$ error, the LoS component will
not be eliminated completely after the projection, and this will pollute the detection of subsequent AoAs.
This limits \name\ to $K=4$ \hl{in practice.}
\minus \minus 

\subsection{When will \name\ fail or suffer?}

\begin{figure}[ht]
    \includegraphics[width=0.95\columnwidth]{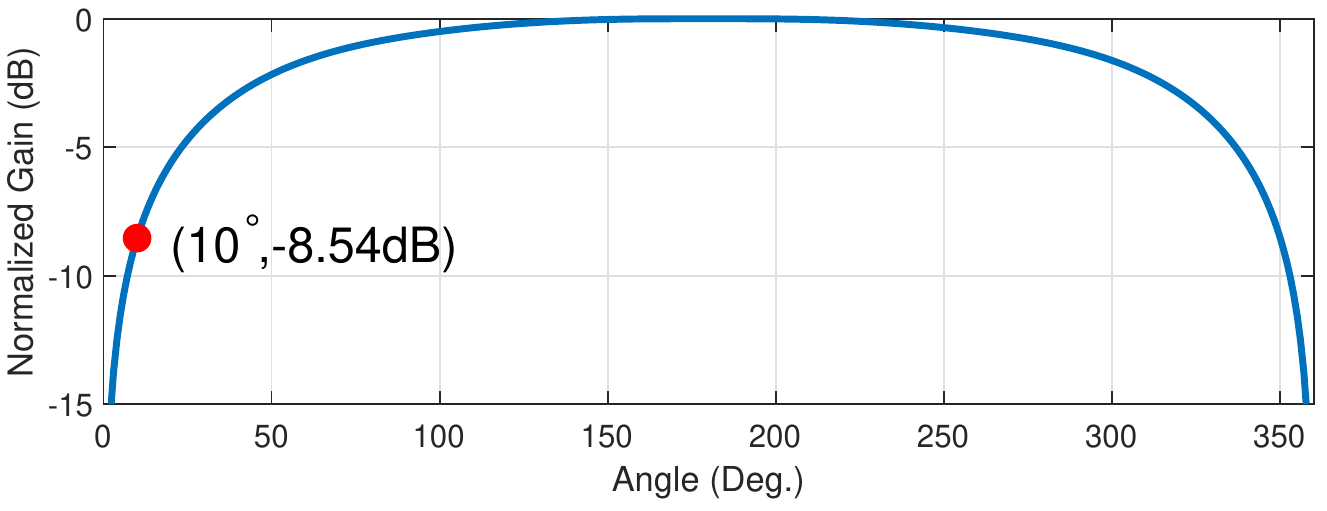}
    \minus
    \caption{Signal attenuation after projection. }
    \label{fig:proj_att}
    \minus \minus 
\end{figure}

While the AoA sub-space offers immunity from correlated signals, it begins to falter when AoAs are angularly close to each other.
For instance, if $\theta_1$ is very close to $\theta_0$, the residual space $\mbf{B}_0$ will be nearly perpendicular to $\mbf{a}_{\theta_1}$.
The echo along $\theta_1$ will suffer a strong attenuation after the projection. 
Meanwhile, if the noise vector is perpendicular to $\mbf{a}_{\theta_0}$, its power will be preserved, and the SNR of the echo will drop proportionally. 
Figure~\ref{fig:proj_att} plots the signal attenuation factor $\mbf{F}_{\theta}$ assuming $\theta_0=0^\circ$. 
If $\theta_1$ is $10^\circ$, the echo suffers an $8.54dB$ attenuation compared to the worst case noise vector (at $180^\circ$).
Of course, the average case is much better.
\new 

\begin{figure}[ht]
    \includegraphics[width=0.49\columnwidth]{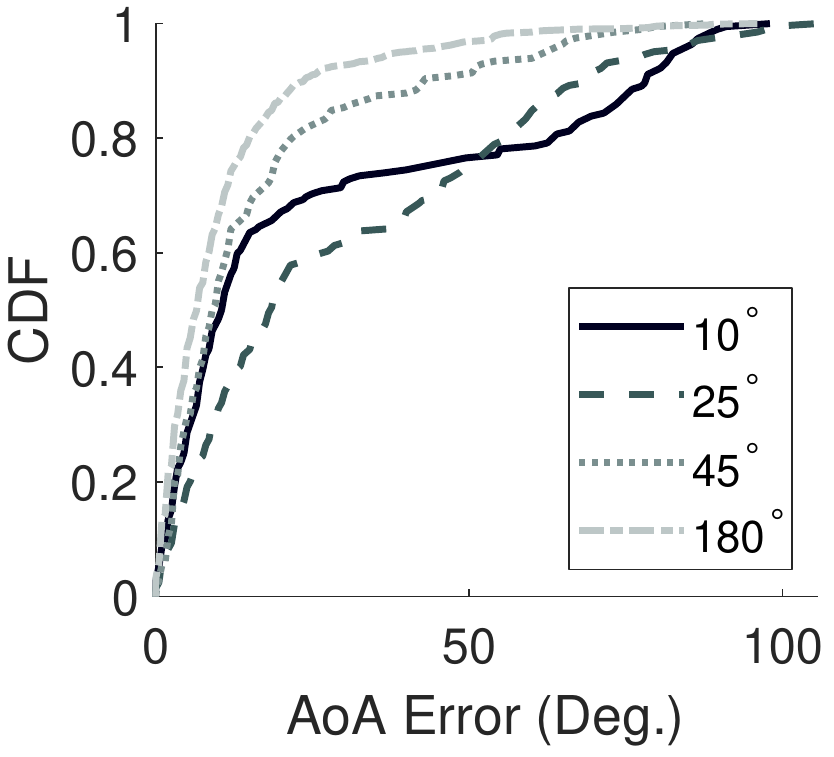}
    \includegraphics[width=0.49\columnwidth]{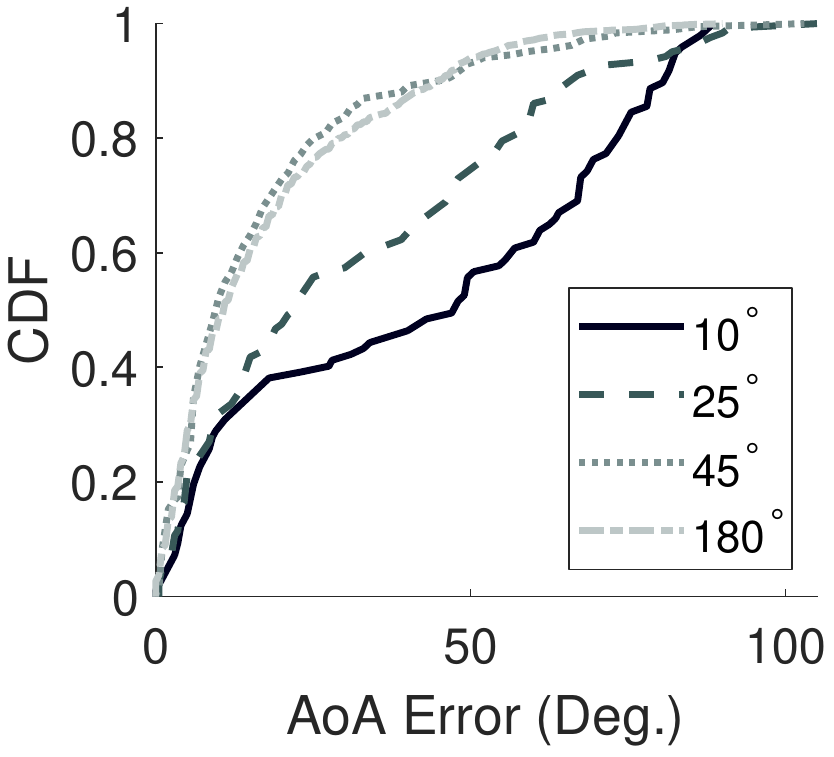}
    \minus
    \caption{Error for (a) $\theta_1$ and (b) $\theta_2$ in different path angular differences. }
    \label{fig:simu_ang_diff}
    \minus
\end{figure}

Figure~\ref{fig:simu_ang_diff} visualizes this shortcoming by plotting the AoA error as a function of the angular difference between the AoA vectors.
The solid black line corresponding to $10^\circ$ indicates all the test cases in which all the AoA's were within $10^\circ$ of each other (i.e., $\theta_0-\theta_k \leq 10$).
Clearly, performance degrades as the AoA vectors come close to each other.
\new




\textbf{Some ideas:} One natural question is: {\em since $F_{\theta_0}$ is known, why not utilize ${F_{\theta_0}}$ as a weighing function to amplify AoAs angularly close to $\theta_0$.}
We explored this idea; however, the problem is that the noise vector is completely unknown.
Thus, it is possible that we would end up multiplying an angularly far away from $\theta_1$ with a small weight, but if the noise vector is close to $\theta_0$, we would amplify the noise.
In other words, solving one problem would create another.
\new

Perhaps one fall-back possibility is to invite VoLoc-style optimization when the AoA sub-space has reduced to few dimensions.
Said differently, when $\theta_0, \theta_1$, and $\theta_2$ have already been detected, and the residual space $B_2$ is small, we can employ optimization to minimize residuals for candidate steering vectors.
This is algorithmically not superior since the accuracy would improve at the cost of computational complexity.
However, if the application permits some time cushion, such hybrid approaches may be practical.

\section{Implementation and Evaluation}

\begin{figure*}[hbt]
    \centering
    \includegraphics[width=0.32\textwidth]{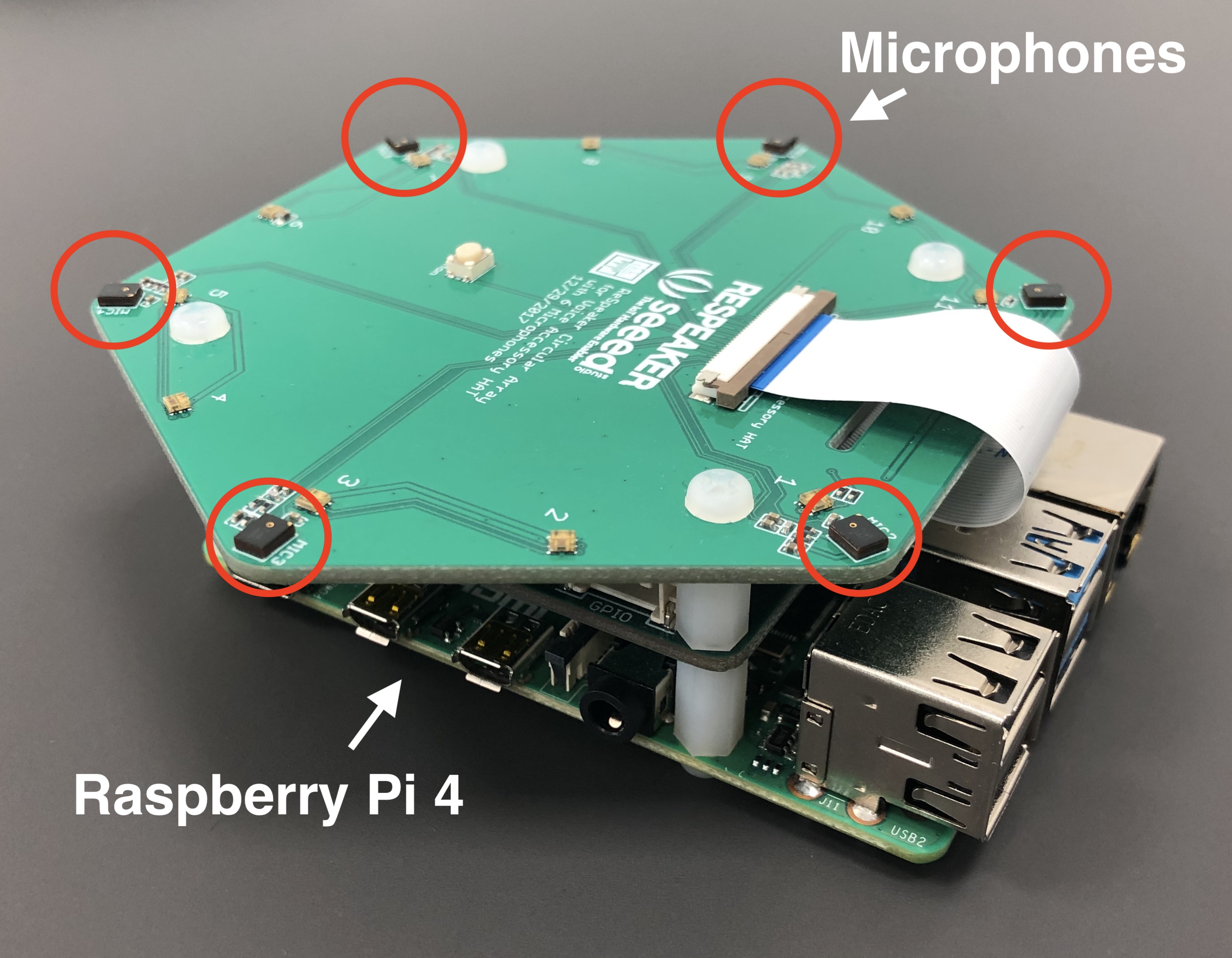}
    \includegraphics[width=0.67\textwidth]{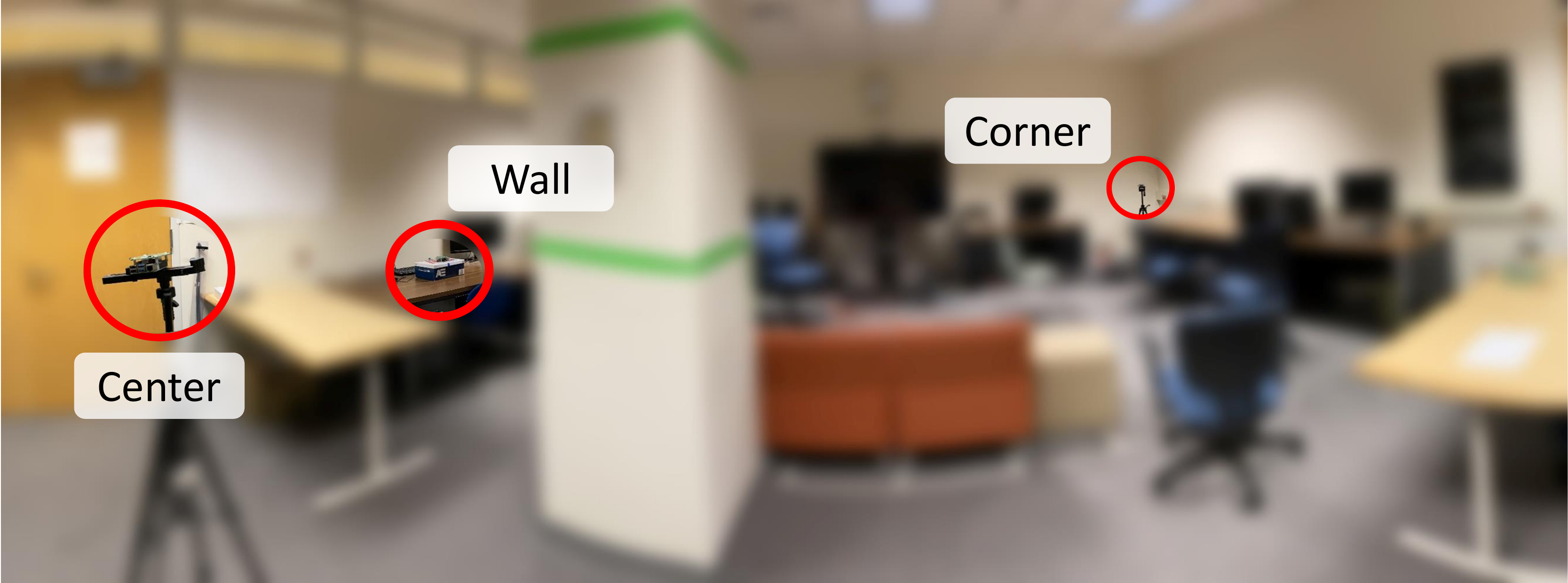}
    \minus \minus
    \caption{(a) The receiver composed of 6-Mic. circular array mounted on Raspberry Pi4. (b) Experiment setup in a lab space, with receivers in $3$ types of AoA environments.}
    \label{fig:respeaker}
    \minus
\end{figure*}

\subsection{Implementation}
\name\ has been implemented on a 6-microphone array from SEEED ~\cite{ReSpeaker}, mounted on a Raspberry Pi 4 ~\cite{RaspberryPi}. 
The microphone arrangement is circular (shown in Figure ~\ref{fig:respeaker}(a)) similar to Amazon Echo ~\cite{Echo}.
We do not use off-the-shelf Amazon/Google smart speakers since they do not expose the raw acoustic samples.
The SEEED platform offers a sampling rate of $16KHz$, covering most of the audible energy spectrum for speech, music, and ambient noise.
We use an iPhone 11 Pro smartphone speaker as the sound source -- we play various types of sounds at $80$\% of the max volume.
\new

Experiments are performed in $2$ multipath settings: (a) a relatively small studio apartment, and (b) a larger engineering lab (Figure ~\ref{fig:respeaker}(b)).
In each setting, we test {\name} in $3$ types of AoA regimes: (a) near the corner, (b) near the side wall, and (c) near the center of the room.
The rooms are fully populated with everyday objects, including furniture, computers, cabinets, TVs, refrigerators, etc.
The iPhone sound source is placed at discrete positions on a circle around the receiver; several concentric circles are formed.
For corners and sides, the circles are limited to quarter circles and half-circles, respectively.
The voice of $3$ volunteers -- one female and two males -- are recorded and played from the iPhone.
The voice signals are composed of various wake-words, including Siri, Google, Bixby, Alexa, and a sentence: "How is the weather today?".
In total, we tested at more than 100 transmitter-receiver location pairs.
\new

To obtain ground truth AoA, we play a known chirp signal spanning $[0,8]KHz$ to estimate the multipath channel at each microphone.
Since each echo creates a peak in the channel, and since the peak from a given echo is time-shifted across microphones, we can carefully derive the AoAs by aligning the peaks between microphone pairs.
Due to ambient and hardware noise, the alignments are not perfect; however, we leverage multiple pairs of microphones to solve a regression on AoA. 
Finally, we actively test our ground-truth technique by placing artificial reflectors at known angles and checking if we are able to detect the expected AoA reliably.
\minus \minus

\subsection{Performance Results}


Experiments are designed to answer the following questions:
\begin{enumerate}[label=(\alph*)]
    \item What is {\name}'s AoA estimation error (compared to ground truth) for the LoS path and subsequent echoes? 
    How does error grow with successive echoes?
    \item How does \name\ compare against existing algorithms, namely MUSIC \cite{MUSIC}, GCC-PHAT \cite{GCC}, and VoLoc \cite{VoLoc}?
    \item How robust is AoA estimation across different distances, noise,  reflection configurations, and users?
    \item What is {\name}'s completion time in comparison to existing AoA algorithms mentioned above? 
\end{enumerate}


\subsection*{Overall AoA Estimation Accuracy}
\begin{figure}[hbt]
    \includegraphics[width=0.95\columnwidth]{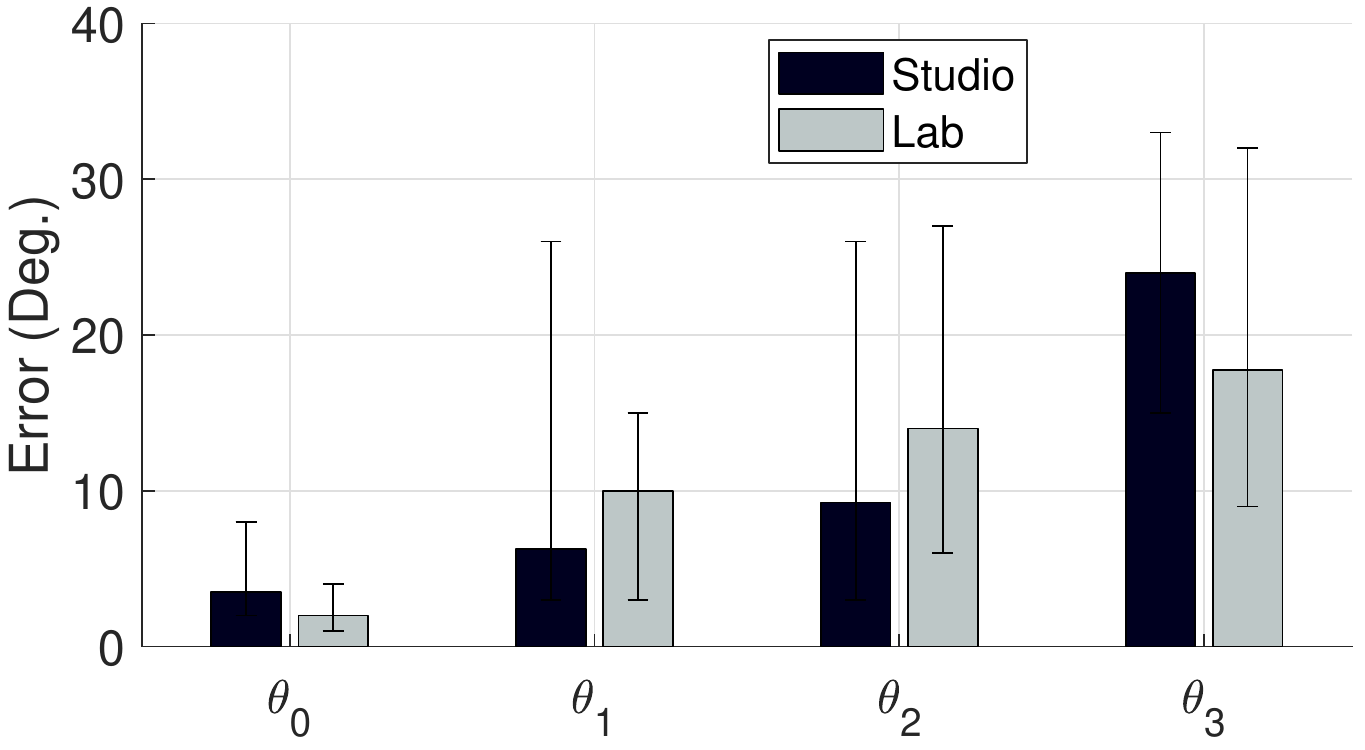}
    \minus
    \caption{{\name}'s median AoA error across all experiments in studio and lab (errorbars: $25$ and $75$ \%ile).}
    \minus \minus 
    \label{fig:env}
    \minus \minus 
\end{figure}

Figure \ref{fig:env} shows {\name}'s median AoA estimation error across all the experiments; the error bars denote the $25^{th}$ and the $75^{th}$ percentile.
On the X-axis, $\theta_0$ represents the AoA of the LoS component, and $\theta_i, i=\{1,2,3\}$ represents the AoA of subsequent echoes.
In the lab setting, the median errors are $2^{\circ}$, $10^{\circ}$,  $14^{\circ}$, and $18^{\circ}$ for $\theta_0$, $\theta_1$, $\theta_2$, and $\theta_3$, respectively. 
The results from the studio are comparable since the size of the room does not matter much.
This is because {\name} is not sensitive to the actual propagation delays for the echoes; only the relative delay at microphones is necessary for AoA.
\new 

\begin{figure}[hbt]
    \includegraphics[width=0.48\columnwidth]{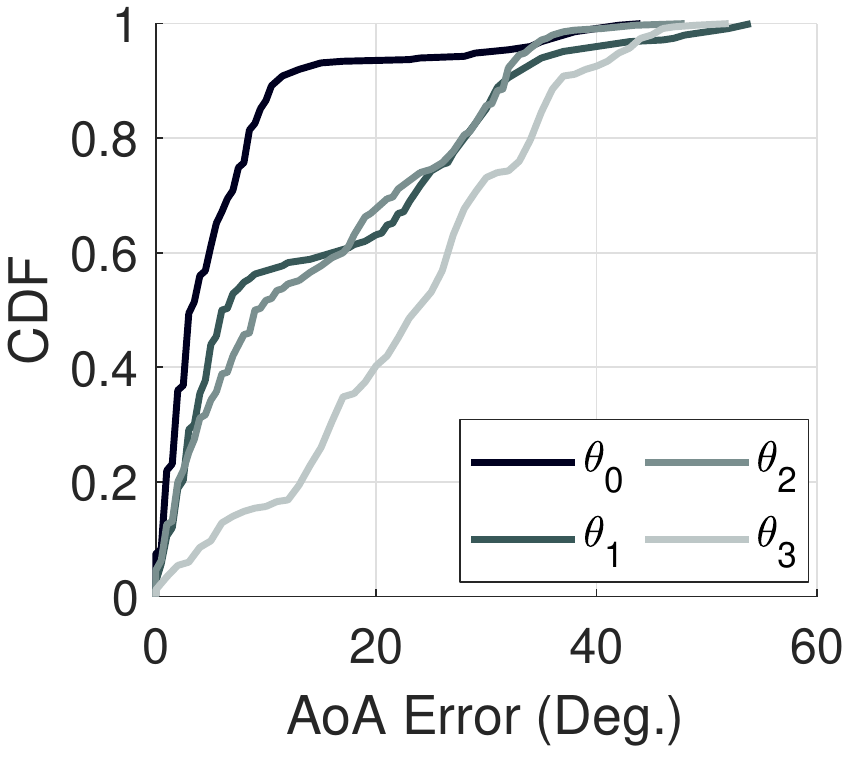}
    \includegraphics[width=0.48\columnwidth]{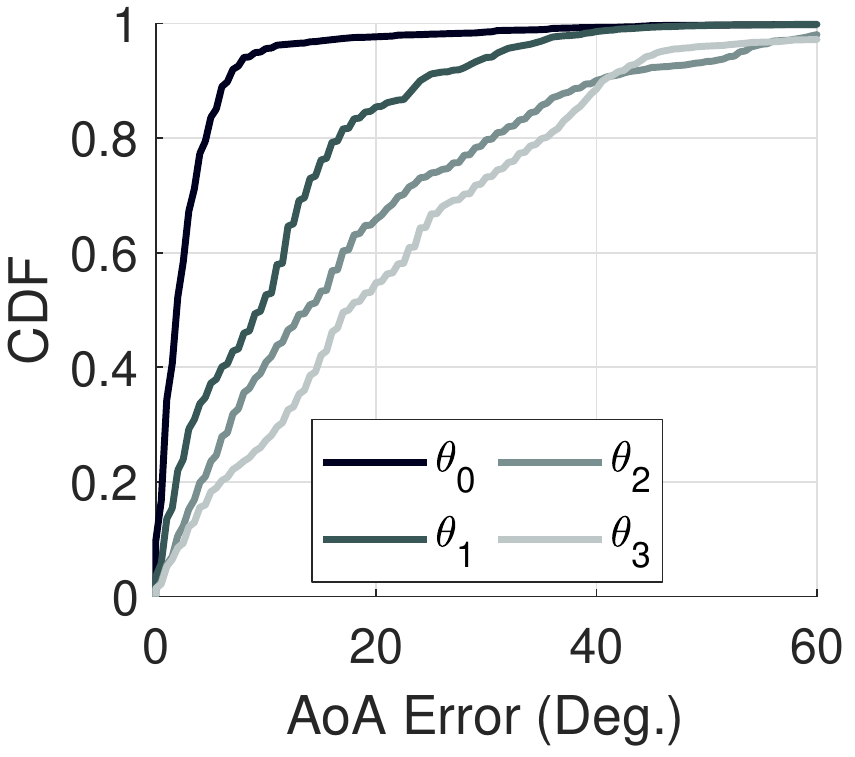}
    \minus
    \caption{CDF plot across all experiments in the (a) small studio and (b) large lab. }
    \label{fig:env_cdf}
    \minus \minus 
\end{figure}

Figure \ref{fig:env_cdf} shows the complete AoA-error CDFs corresponding to the above results.
Beyond $K=4$ AoAs, the median AoA error grows more than $25^\circ$, either because the signals are considerably weaker, or in some cases,  co-aligned with the earlier AoAs.

\subsection*{Comparison with Existing Algorithms}
\begin{figure}[ht]
    \includegraphics[width=0.95\columnwidth]{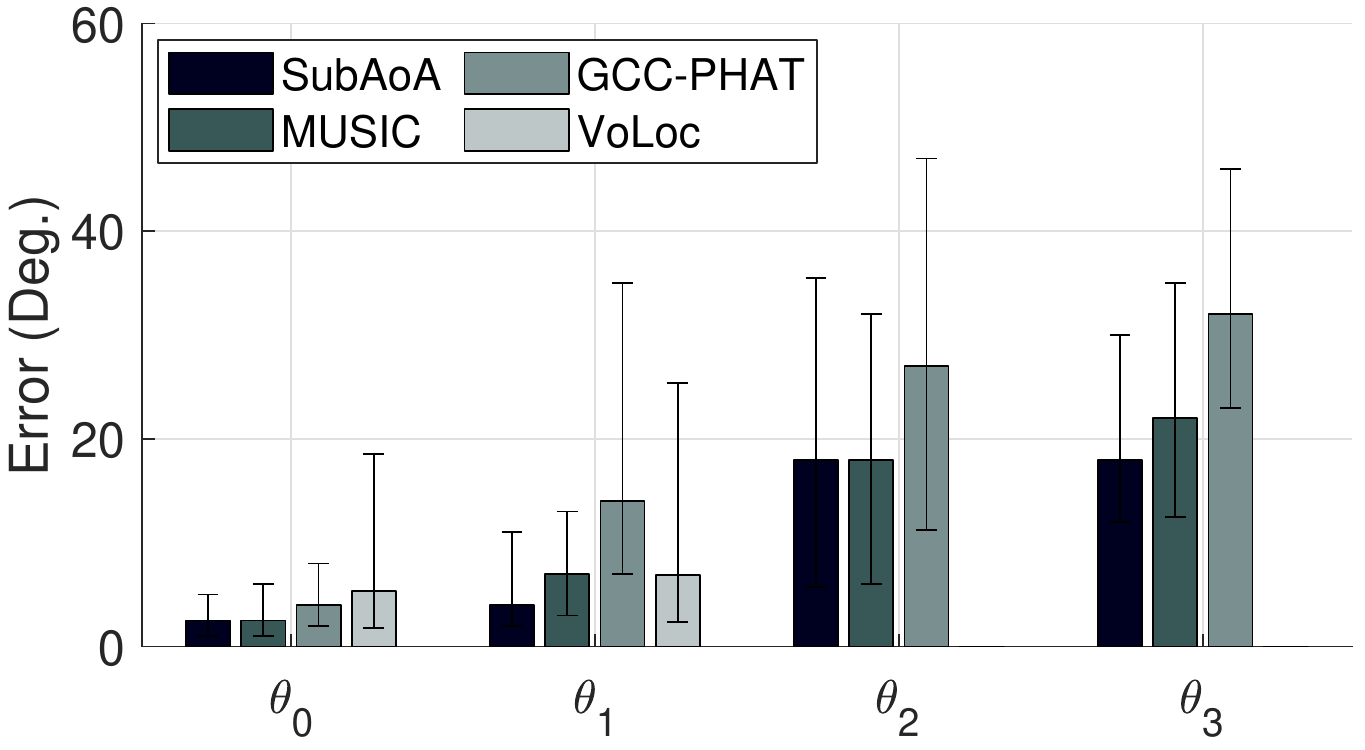}
    \includegraphics[width=0.95\columnwidth]{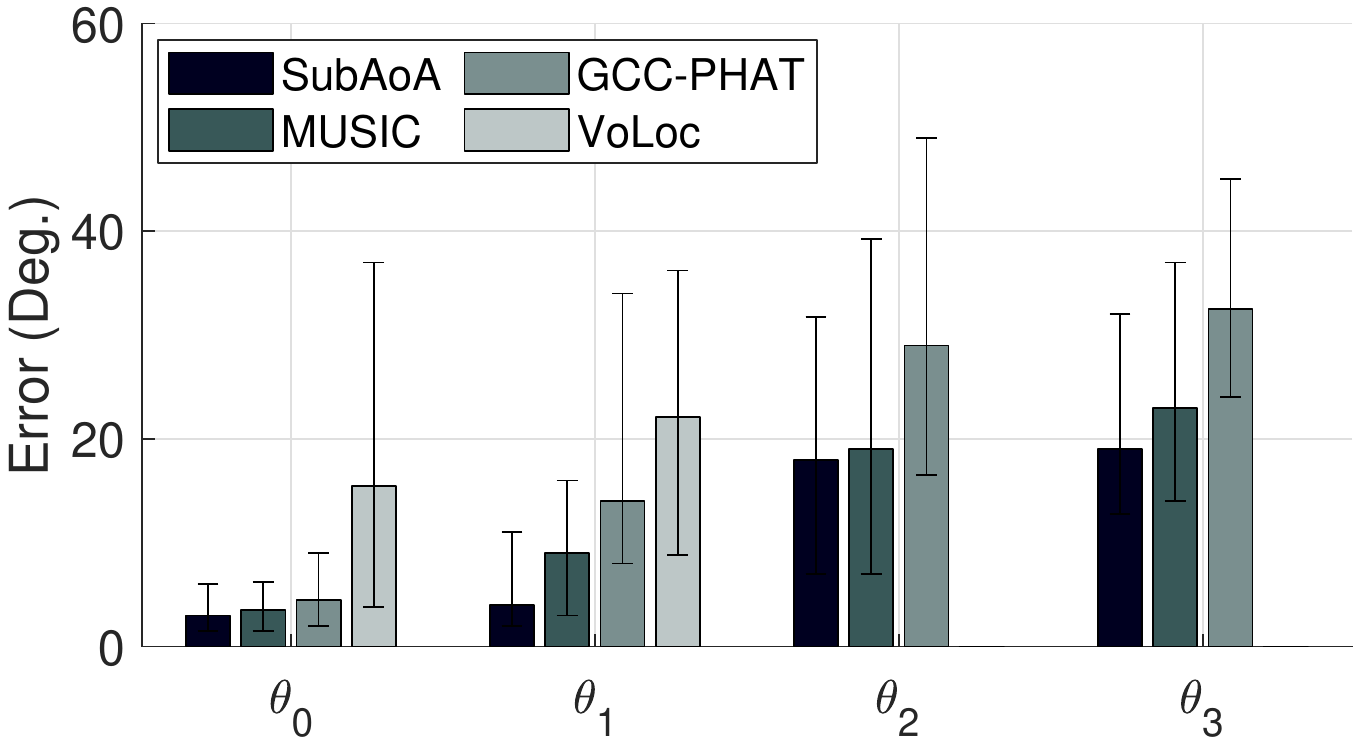}
    \minus
    \caption{Performance comparison of {\name}, MUSIC, GCC-PHAT, and \hl{VoLoc} under (a) quiet (b) noisy (10dB SNR) environment.}
    \label{fig:alg}
    \minus \minus 
\end{figure}

Figure \ref{fig:alg} compares of {\name}'s performance with $3$ existing  algorithms: (a) MUSIC~\cite{MUSIC},  (b) GCC-PHAT \cite{GCC}, and (c) VoLoc~\cite{VoLoc} in both the quiet and noisy environment.
We run MUSIC with $K=4$ uncorrelated source signals, and pick the $\theta_{\{0:3\}}$ AoAs with the highest likelihood. 
For GCC-PHAT, we pick the $\theta_{\{0:3\}}$ with highest normalized correlation coefficient. 
Finally, since VoLoc needs a nearby wall reflector, we \wally{place the receiver near the wall, and} assume the distance to the wall is perfectly known (this is clearly favorable to VoLoc). 
Also, we note that VoLoc only outputs $\theta_{\{0:1\}}$ because the complexity of computing $\theta_2$ is prohibitively high.
\new 


\wally{For the quiet setting,} compared to GCC-PHAT's median errors of $14^{\circ}$ and $27^{\circ}$ (for $\theta_1$ and $\theta_2$), \name\ reduces the errors by $71\%$ and  $33\%$ respectively. 
Beyond $\theta_2$, GCC-PHAT's correlation approach begins to derail due to indoor multipath; as a result, the \hl{error bars grow large.}
\hl{VoLoc performs well for} $\theta_0$ and $\theta_1$,  but \name\ is still better.
But for later AoAs, and in terms of running time, VoLoc is certainly inferior to {\name}. 
MUSIC's sub-space approach balances multiple objectives well -- it detects $K=4$ AoAs and runs in real time.
However, {\name} still outperforms MUSIC since the strong correlation in multipath signals affects the latter.
In noisy environments, VoLoc fails in estimating even the LoS AoA because it cannot correctly find a clean start of the speech (a necessary assumption), and gaussian noises heavily affect the cancellation process.
\new

\begin{figure}[ht]
    \includegraphics[width=0.95\columnwidth]{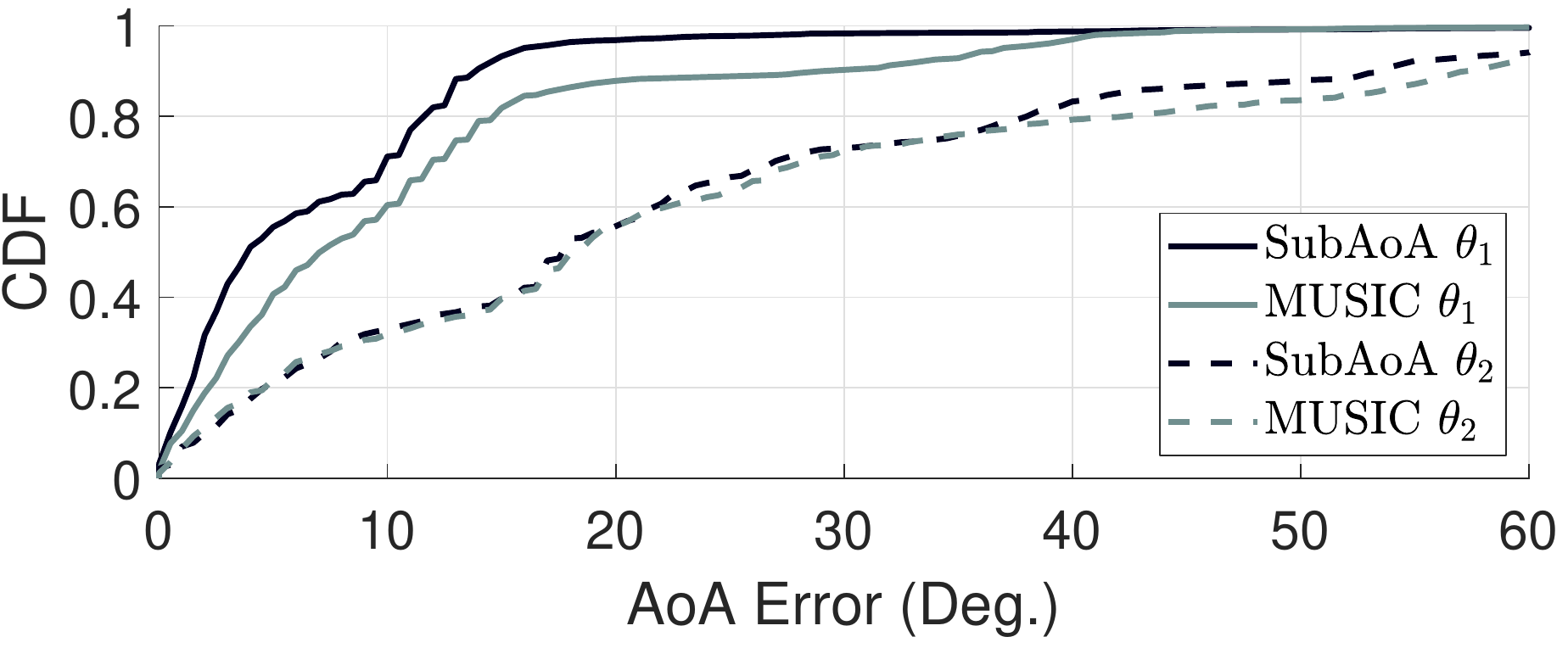}
    \includegraphics[width=0.95\columnwidth]{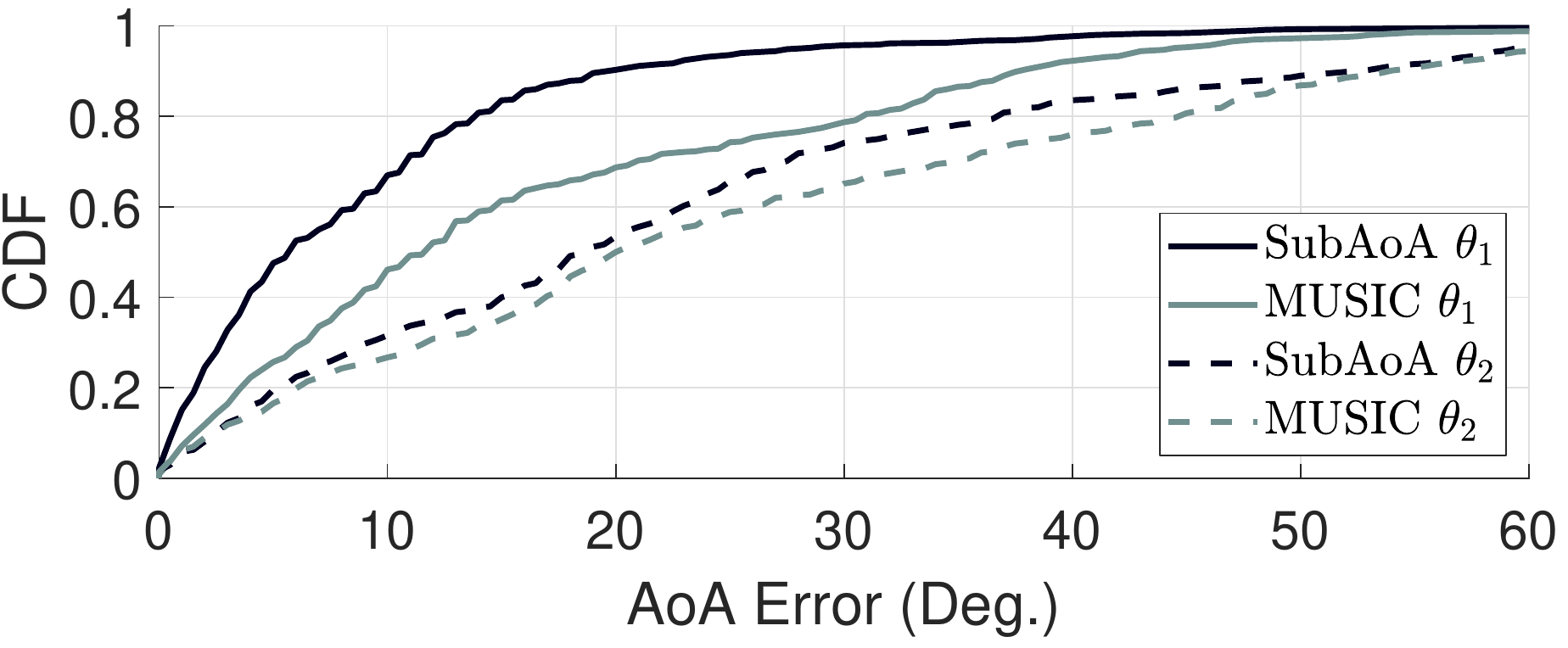}
    \minus
    \caption{CDF plot of $\theta_2$ and $\theta_3$ estimation error for different algorithms when the SNR is (a) 20dB (b) 0dB.}
    \label{fig:music}
    \minus \minus 
\end{figure}

\begin{figure}[ht]
    \includegraphics[width=0.46\columnwidth]{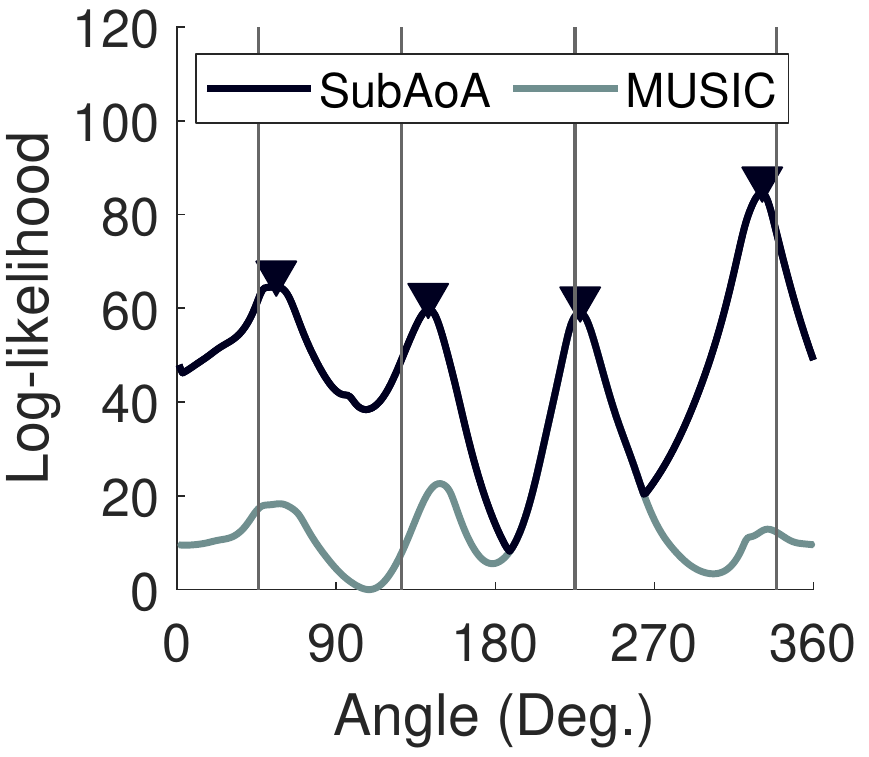}
    \includegraphics[width=0.46\columnwidth]{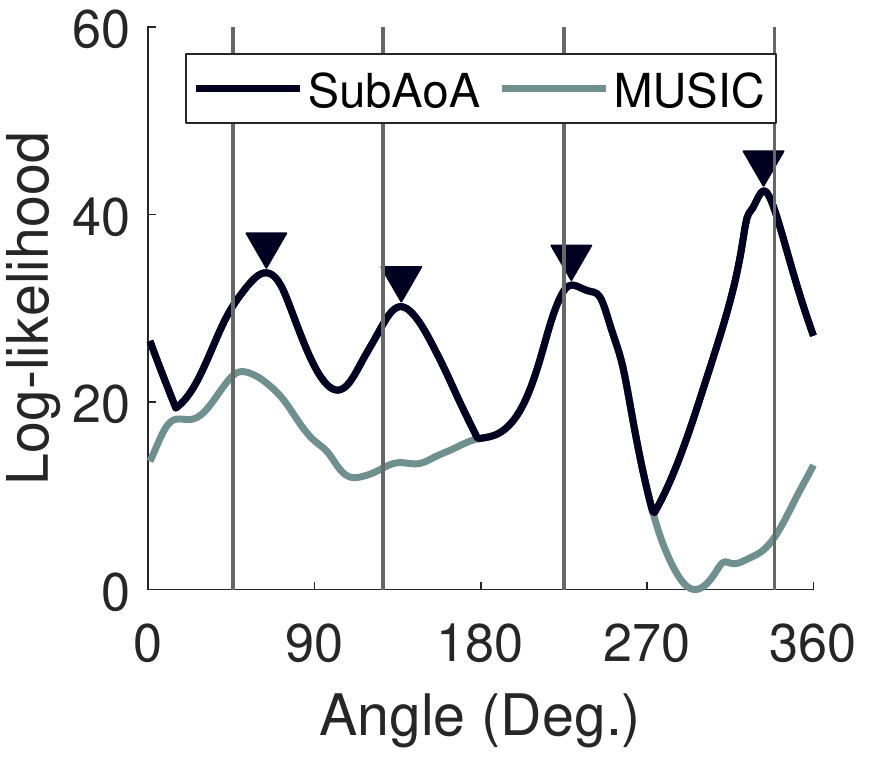}
    \minus
    \caption{AoA spectrum of \name\ and MUSIC when the SNR is (a) 20dB (b) 0dB.
    The black lines mark the ground truth AoAs.}
    \label{fig:spec}
    \minus \minus 
\end{figure}

Figure \ref{fig:music} shows the performance breakdown in two SNR regimes, $20$dB and $0$dB (i.e., the white gaussian background noise is increased for both MUSIC and {\name}).
In low SNR regimes, MUSIC suffers more because the \hl{signal sub-space is directly affected, while the AoA sub-space is less sensitive.} 
Compared to MUSIC, \name\ reduces $\theta_1$'s median error by $47\%$ in the $20dB$ regime, and $50\%$ when the SNR is $0dB$.
Figure \ref{fig:spec} zooms into the AoA spectrum of \name\ and MUSIC in a multi-echo environment.
In MUSIC, we can clearly see that the weak AoA components are buried in the noise when the SNR decreases, while \name\ maintains consistently strong peak heights for $\theta_{0:3}$.
\minus


\subsection*{Impact of Distance and SNR}
\begin{figure}[hbt]
    \includegraphics[width=0.95\columnwidth]{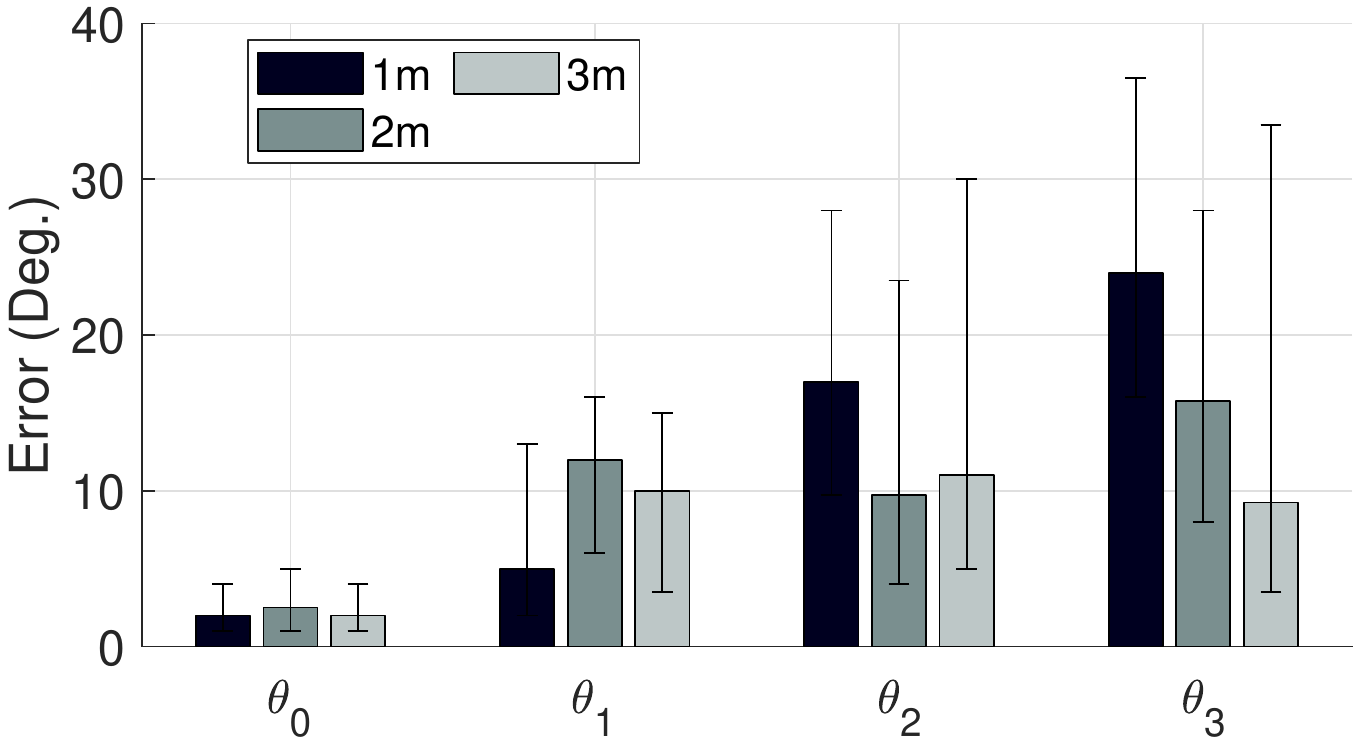}
    \minus \minus 
    \caption{{\name}'s error vs. transmission distances.}
    \label{fig:dist}
    \minus \minus 
\end{figure}

Figure \ref{fig:dist} shows {\name}'s performance with increasing distance between the speaker and receiver.
Evidently, there is no obvious impact of distance.
This is because (1) distance impacts SNR less than the energy absorption due to reflections, and (2) {\name} is not heavily sensitive to SNR since the decoding is in the AoA sub-space.
Thus while the error grows with distance for $\theta_1$, it somewhat reduces for $\theta_3$.
\new

\begin{figure}[hbt]
    \includegraphics[width=0.95\columnwidth]{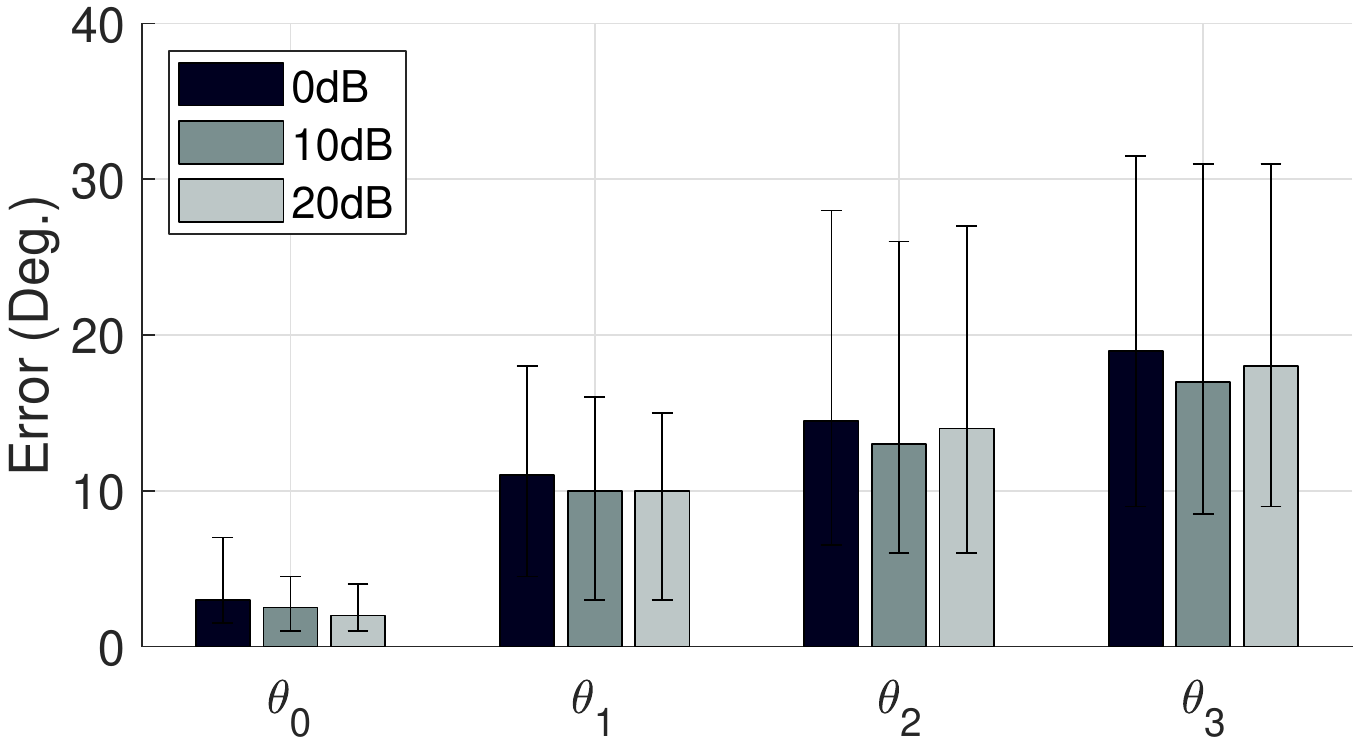}
    \minus \minus 
    \caption{\name\'s error across different SNR levels.}
    \label{fig:noise}
    \minus \minus 
\end{figure}

With COVID, the background noises were particularly low in the lab settings.
To evaluate how background noise would affect {\name}'s performance (i.e., at lower SNR), we add white gaussian noises to the received signal. 
Figure ~\ref{fig:noise} shows the error across different SNR levels. 
Evidently, performance hardly degrades with the SNR level. 
This provides stronger evidence to {\name}'s robustness to SNR, and hence, the value of operating in the AoA sub-space.
\new

\subsection*{Impact of Rx Location (Corner, Side, Center)}
\begin{figure}[ht]
    \includegraphics[width=0.95\columnwidth]{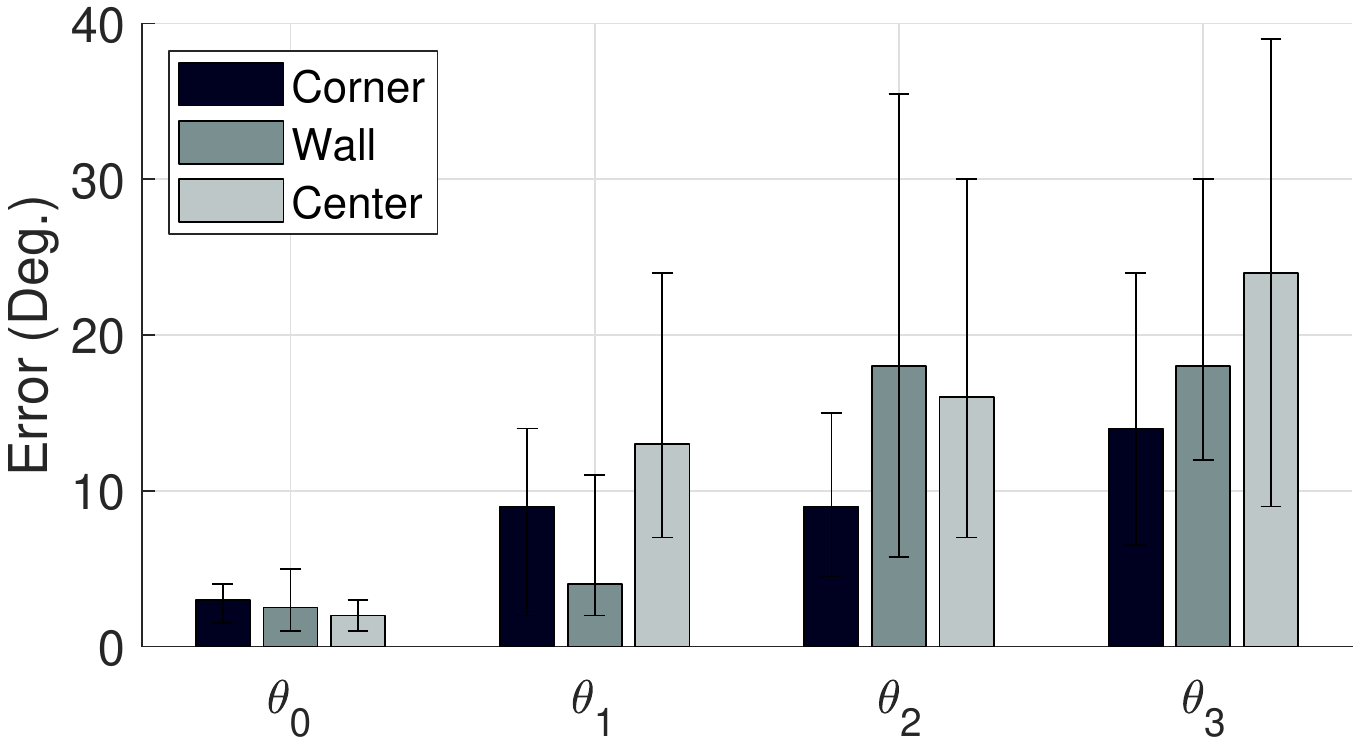}
    \minus
    \caption{Impact of receiver placed in different AoA configurations (corner, side, center).}
    \label{fig:regime}
    \minus
\end{figure}

\begin{figure}[ht]
    \includegraphics[width=0.4\columnwidth]{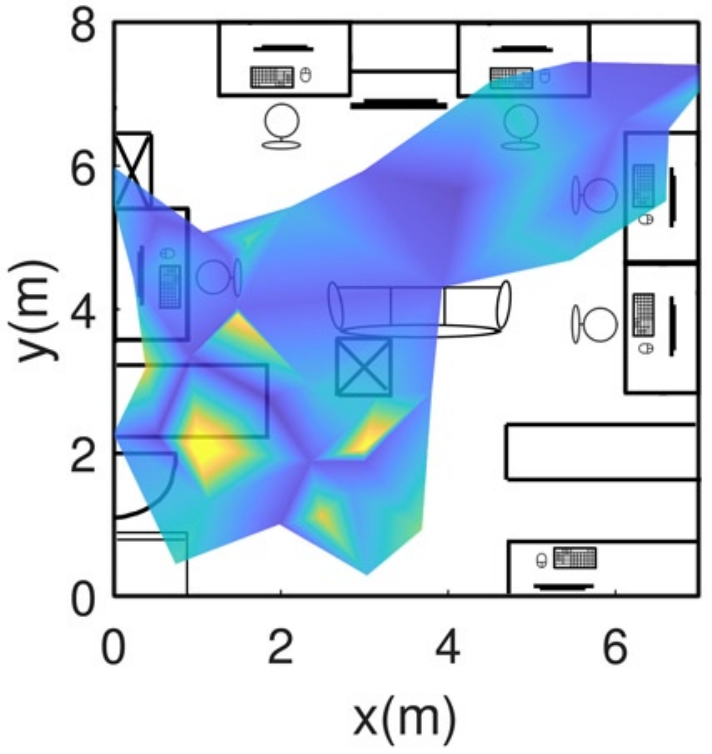}
    \includegraphics[width=0.57\columnwidth]{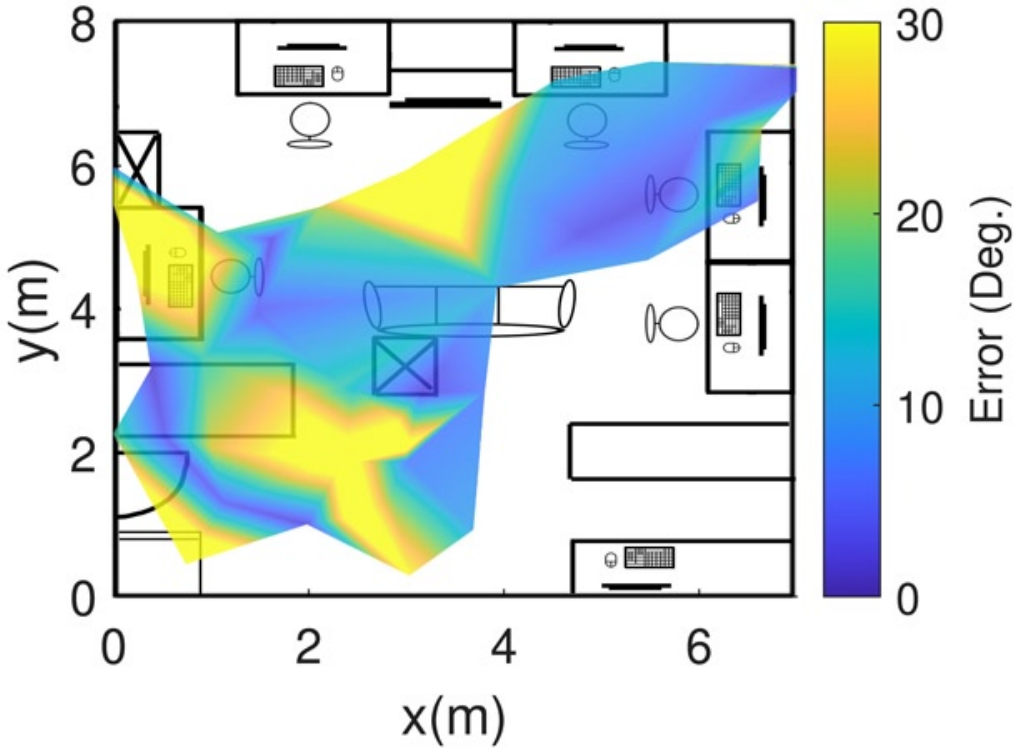}
    \minus
    \caption{Median error of (a) $\theta_1$ (b) $\theta_2$ across different speaker and device locations}
    \label{fig:heat}
    \minus \minus \minus \minus 
\end{figure}

We test {\name}'s performance under $3$ AoA environments, i.e., placed near the corner (2 nearby walls), side (1 nearby wall), and center of a room (no nearby walls).
These configurations influence the number of AoAs, their angular separations, and strengths.
Figure ~\ref{fig:regime} plots the median AoA errors.
For the corner setting, the first and second echoes from the walls are strong (and angularly well separated) so the errors are least.
For the single-wall setting, because echoes from the environment are weaker than the wall reflection, the $\theta_2$ error jumps up compared to $\theta_1$.
\hl{When the device is placed at the center of the room, all the echoes are from the far-away walls}, resulting in higher $\theta_{1:3}$ errors.
Figure ~\ref{fig:heat} visualizes the AoA errors for $\theta_1$ and $\theta_2$ in the lab settings.
This heatmap visualizes and confirms that errors are higher near the centers and least near corners or with various reflectors nearby. 
\minus \minus 

\subsection*{Robustness to Source Signal}
\begin{figure}[ht]
    \includegraphics[width=0.95\columnwidth]{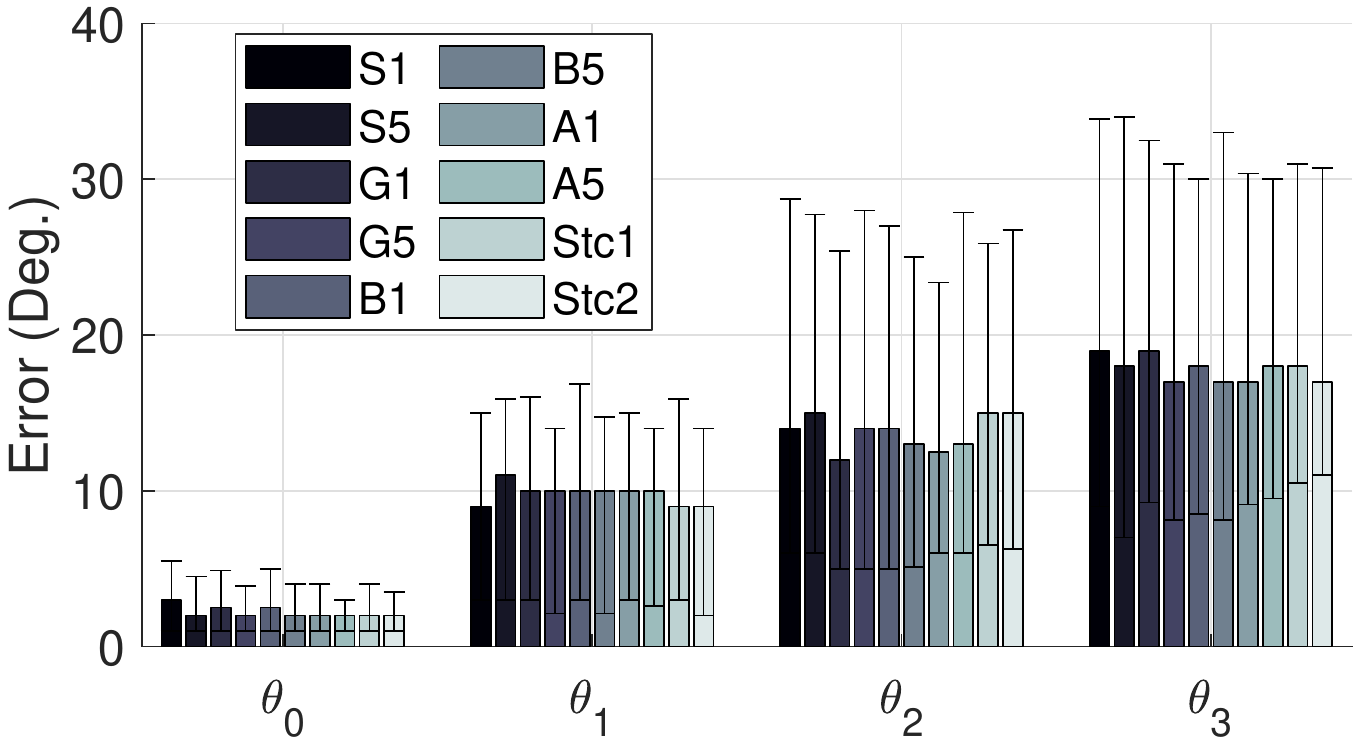}
    \minus
    \caption{{\name}'s error across different speech content. The errorbars denote the $25$ and $75$ percentiles.}
    \label{fig:word}
    \minus
\end{figure}

Figure \ref{fig:word} verifies that the performance of \name\ is not sensitive to the source signal, such as different users or the speech signals they produce.
We play different speech waveforms: S="Siri", G="Google", B="Bixby", A="Alexa", and a sentence Stc="How is the weather today".
We also repeat the words to aid AoA detection with a longer waveform.
Hence, S5 indicates "Siri" repeated $5$ times.
\new 

Evidently, the source (speech) signal does not affect the AoA accuracy much; 
This is not surprising because the techniques underlying \name\ does not utilize any structures of the signal, like base frequency, harmonics, pauses, correlation, etc.
Longer speech (i.e., sentences or repeats) help slightly because it suppresses the error from random noise. 
\new 

\begin{figure}[ht]
\minus \minus 
    \includegraphics[width=0.95\columnwidth]{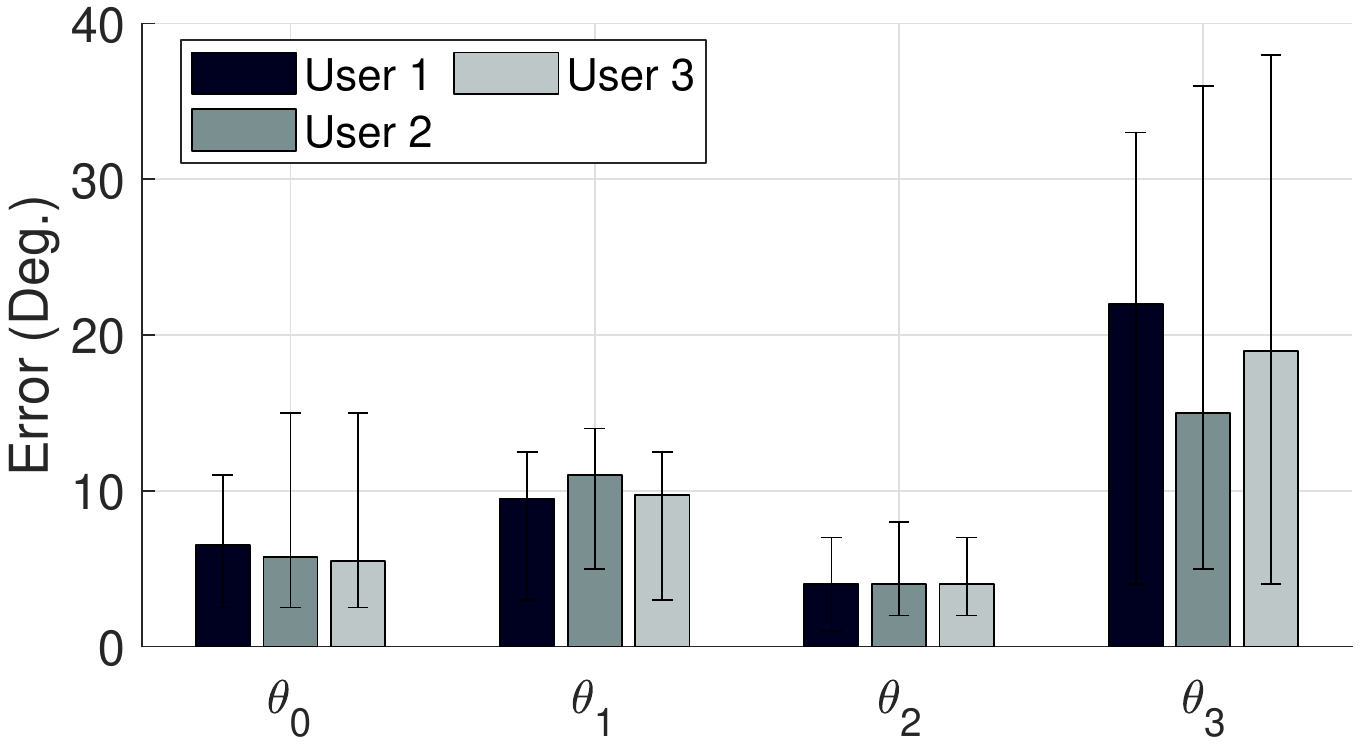}
    \minus
    \caption{{\name}'s error across different user.}
    \label{fig:user}
    \minus
\end{figure}

To test {\name}'s sensitivity to different users (speaking the same words), we record the speech from $3$ volunteers and play them from identical locations. 
Figure \ref{fig:user} shows the results. 
The median error of $\theta_1$, for example, are $9.5^\circ$, $11^\circ$, and $9.75^\circ$ across the users. 
Again, these results confirm that {\name} is robust, hence generalizable to any voice signal.
\new 

\subsection*{Computation Time}
We evaluate \name\ on a laptop equipped with AMD Ryzen 7 4800H, 2.9GHz, and 32GB Memory. 
Table ~\ref{tbl:computation} shows the median computation time of different algorithms.
The overhead of \name\ is similar to MUSIC and GCC-PHAT, but much lower than VoLoc.
This is because VoLoc needs to solve a minimization problem on path delays and amplitudes (in order to perform accurate cancellation).
{\name} side-steps that by estimating null-spaces for the decoded AoA.
Observe that even though {\name} takes higher compute time than MUSIC, they both are sub-seconds, hence can support real-time applications in the acoustic time-scale.
\new

\begin{table}[t]
    \centering
    \begin{tabular}{|c|c|c|c|c|}
    \hline
    Algorithm & \name\ & MUSIC & GCC-PHAT & VoLoc \\\hline 
    Median time (ms) & 537.7 & 212.2 & 423.3& 6672.0\\\hline
    \end{tabular}
    \caption{Median computation time in milliseconds across different algorithms}
    \minus \minus \minus \minus \minus \minus 
    \label{tbl:computation}
\end{table}

\section{Limitations and Next Steps}

\textbf{Multi-AoA for RF Signals} \\ 
Although this paper has been presented and evaluated with acoustic signals, the core \name\ algorithm is  generalizable to any signals, including RF.
In no part of the algorithm are there assumptions around acoustic/speech source-signals; neither are there implicit assumptions about the slow nature of sounds.
We have chosen acoustics here since human voice/speech is an important application where the source signal is unknown.
We leave the treatment and evaluation with RF signals to future work.
\new 


\textbf{From AoA to (Blind) Channel Estimation} \\ 
Another opportunity lies in approaching blind channel inference (BCI) through AoA.
In other words, while \name\ is extracting relative delays between microphones, the only unknowns that remain are the absolute path delays and amplitude attenuation.
Previous BCI algorithms like JADE~\cite{JADE} estimate the path delays and AoAs jointly, resulting in very high computational complexity. 
With \name\ estimating the AoAs accurately, the problem of path delays/amplitudes seems reachable.
We believe \name\ ushers in such ideas.




\section{Additional Related Work}
\textbf{$\blacksquare$ Acoustic AoA Estimation:} 
AoA estimation on microphone arrays is clearly a rich area. 
Many papers extend the foundational ideas from Section~\ref{sec:primer} to different settings ~\cite{ESPRIT, JADE, xu2017weighted, benesty2000adaptive, freire2011doa, benesty2004time, svaizer2012environment}.
ESPRIT~\cite{ESPRIT} clusters microphones into different groups to reduce the complexity of MUSIC.
JADE ~\cite{JADE} targets multipath scenarios and jointly optimizes the path AoA and delay from the received signals (with similarities to blind estimation ~\cite{micArrayBCI,tongBCISurvey98,tong1999joint,hua1996fast,lin2008blind, lin2007blind}.
Newer ideas introduce deep-learning/training phase to infer rich channel information ~\cite{nakadai2005sound,nakajima2009real,niwa2010estimation,brutti2011inference}.
Nakadai et al.~\cite{nakadai2005sound} uses a large microphone to capture and match the speaker directionality pattern. 
Brutti et al.~\cite{brutti2011inference} models the low-order reflection and accounts for the directional radiation pattern with modified GCC-PHAT algorithm.
Compared to these algorithms, \name\ shows a more foundational opportunity in the AoA sub-space representation.
Perhaps many other works can exploit the AoA sub-space.
\new 

\textbf{$\blacksquare$ RF AoA Estimation:} 
Radio frequency (RF) AoA estimation is another important and thriving topic ~\cite{sit2012direction,bialer2019performance,karanam2018magnitude,haider2018listeer,ghasempour2020single,oumar2012comparison,xie2019md,180296,Kotaru:2015:SDL:2785956.2787487}. 
Knowing path AoA enables beamforming techniques in RF communication (especially mmWave).
Karanam et al. ~\cite{karanam2018magnitude} utilizes only the received amplitude at each antenna to estimate the AoA.
In ~\cite{bialer2019performance}, a deep learning technique is introduced to estimate a large number AoAs accurately. 
Ghasempour et al. ~\cite{ghasempour2020single} creatively utilizes the leaky-wave devices' radiation characteristics to detect both the AoA and Angle of Departure (AoD) in a single-shot measurement.
{\name} applies to RF signals as well, and could complement new types of antennas as proposed by ~\cite{ghasempour2020single}.
Finally, {\name} could find critical applications in mmWave beamforming, mobility tracking, and link re-establishments. 
\new 

\textbf{$\blacksquare$ Space of Applications:} 
AoA estimation is also gaining prominence in personal computing and IoT.
Vast literature ~\cite{yun2017strata,VoLoc,sallai2011weapon,Lazik:2015:DAA,dokmanic2013acoustic,rossi2013roomsense} explores a number of acoustic sensing capabilities.
Yun et al.~\cite{yun2017strata} uses the signals from a smartphone to track human fingers, providing a device-free HCI interface for VR/AR experience.
In ~\cite{dokmanic2013acoustic}, the authors transmit a sweeping sine wave to estimate the multipath and compute the room geometry from the reverberations.
~\cite{rossi2013roomsense} develops an indoor positioning system based on acoustic active fingerprinting using a phone. 
Finally, UK and EU have passed legislation \cite{uk-car-noise, bbc-car-noise} that electric cars broadcast artificial sounds for the safety of pedestrians and other vehicles; this implies that a car might be able to hear other cars around the corner (even if they cannot see them through cameras and LIDARs). 
\name\ can serve as a foundational building block for these and many other applications.
\minus \minus 



\section{Conclusion}
\label{sec:remarks}
We develop {\name}, an algorithm that factors out multiple AoAs $\theta_{0:K-1}$ from a microphone array.
The technique shows the advantages of operating in the AoA sub-space, instead of the conventional signal sub-space.
We observe that the algorithm's behavior is amenable to practical settings, including unknown source signals, correlated multi-path, ambient noise, etc.
We believe \name\ could usher additional ideas in the algorithm as well as the space of applications.




\bibliographystyle{IEEE}
\bibliography{reference}

\begin{thebibliography}{10}

\bibitem{dibiase2000high}
Joseph~Hector DiBiase,
\newblock {\em A high-accuracy, low-latency technique for talker localization
  in reverberant environments using microphone arrays},
\newblock Brown University Providence, RI, 2000.

\bibitem{VoLoc}
Sheng Shen, Daguan Chen, Yu-Lin Wei, Zhijian Yang, and Romit~Roy Choudhury,
\newblock ``Voice localization using nearby wall reflections,''
\newblock in {\em Proceedings of the 26th Annual International Conference on
  Mobile Computing and Networking}, 2020, pp. 1--14.

\bibitem{262015}
Mei Wang, Wei Sun, and Lili Qiu,
\newblock ``{MAVL}: Multiresolution analysis of voice localization,''
\newblock in {\em 18th {USENIX} Symposium on Networked Systems Design and
  Implementation ({NSDI} 21)}, Boston, MA, Apr. 2021, {USENIX} Association.

\bibitem{autocar_acoustic1}
EA~King, A~Tatoglu, D~Iglesias, and A~Matriss,
\newblock ``Audio-visual based non-line-of-sight sound source localization: A
  feasibility study,''
\newblock {\em Applied Acoustics}, p. 107674, 2020.

\bibitem{uk-car-noise}
``Automated and electric vehicles bill (15th november 2017),'' 2021.

\bibitem{bbc-car-noise}
``{}electric cars: New vehicles to emit noise to aid safety,'' .

\bibitem{kim2006blind}
Taesu Kim, Hagai~T Attias, Soo-Young Lee, and Te-Won Lee,
\newblock ``Blind source separation exploiting higher-order frequency
  dependencies,''
\newblock {\em IEEE transactions on audio, speech, and language processing},
  vol. 15, no. 1, pp. 70--79, 2006.

\bibitem{li2011bayesian}
Xue Li, Steven Hong, Zhu Han, and Zhiqiang Wu,
\newblock ``Bayesian compressed sensing based dynamic joint spectrum sensing
  and primary user localization for dynamic spectrum access,''
\newblock in {\em 2011 IEEE Global Telecommunications Conference-GLOBECOM
  2011}. IEEE, 2011, pp. 1--5.

\bibitem{joshi2013pinpoint}
Kiran Joshi, Steven Hong, and Sachin Katti,
\newblock ``Pinpoint: Localizing interfering radios,''
\newblock in {\em 10th $\{$USENIX$\}$ Symposium on Networked Systems Design and
  Implementation ($\{$NSDI$\}$ 13)}, 2013, pp. 241--253.

\bibitem{GCC}
Charles Knapp and Glifford Carter,
\newblock ``The generalized correlation method for estimation of time delay,''
\newblock {\em IEEE transactions on acoustics, speech, and signal processing},
  vol. 24, no. 4, pp. 320--327, 1976.

\bibitem{MUSIC}
R.~Schmidt,
\newblock ``Multiple emitter location and signal parameter estimation,''
\newblock {\em IEEE Transactions on Antennas and Propagation}, vol. 34, no. 3,
  pp. 276--280, March 1986.

\bibitem{tong1999joint}
Lang Tong and Qing Zhao,
\newblock ``Joint order detection and blind channel estimation by least squares
  smoothing,''
\newblock {\em IEEE Transactions on Signal Processing}, vol. 47, no. 9, pp.
  2345--2355, 1999.

\bibitem{ESPRIT}
Arogyaswami Paulraj, Richard Roy, and Thomas Kailath,
\newblock ``Estimation of signal parameters via rotational invariance
  techniques-esprit,''
\newblock in {\em Nineteeth Asilomar Conference on Circuits, Systems and
  Computers, 1985.} IEEE, 1985, pp. 83--89.

\bibitem{benesty2000adaptive}
Jacob Benesty,
\newblock ``Adaptive eigenvalue decomposition algorithm for passive acoustic
  source localization,''
\newblock {\em the Journal of the Acoustical Society of America}, vol. 107, no.
  1, pp. 384--391, 2000.

\bibitem{JADE}
Michaela~C Vanderveen, Constantinos~B Papadias, and Arogyaswami Paulraj,
\newblock ``Joint angle and delay estimation ({JADE}) for multipath signals
  arriving at an antenna array,''
\newblock {\em IEEE Communications letters}, vol. 1, no. 1, pp. 12--14, 1997.

\bibitem{capon1969high}
Jack Capon,
\newblock ``High-resolution frequency-wavenumber spectrum analysis,''
\newblock {\em Proceedings of the IEEE}, vol. 57, no. 8, pp. 1408--1418, 1969.

\bibitem{manocha_2018}
Inkyu An, Myungbae Son, Dinesh Manocha, and Sung-Eui Yoon,
\newblock ``Reflection-aware sound source localization,''
\newblock {\em 2018 IEEE International Conference on Robotics and Automation
  (ICRA)}, May 2018.

\bibitem{showen2009acoustic}
Robert~L Showen, Robert~B Calhoun, and Jason~W Dunham,
\newblock ``Acoustic location of gunshots using combined angle of arrival and
  time of arrival measurements,'' 2009,
\newblock US Patent 7,474,589.

\bibitem{tongBCISurvey98}
Lang Tong and Sylvie Perreau,
\newblock ``Multichannel blind identification: From subspace to maximum
  likelihood methods,''
\newblock {\em Proceedings of the IEEE}, vol. 86, no. 10, pp. 1951--1968, 1998.

\bibitem{li2016norm}
Yingsong Li, Yanyan Wang, and Tao Jiang,
\newblock ``Norm-adaption penalized least mean square/fourth algorithm for
  sparse channel estimation,''
\newblock {\em Signal processing}, vol. 128, pp. 243--251, 2016.

\bibitem{tugnait2003channel}
Jitendra~K Tugnait and Weilin Luo,
\newblock ``On channel estimation using superimposed training and first-order
  statistics,''
\newblock in {\em 2003 IEEE International Conference on Acoustics, Speech, and
  Signal Processing, 2003. Proceedings.(ICASSP'03).} IEEE, 2003, vol.~4, pp.
  IV--624.

\bibitem{xu2017weighted}
Chenglin Xu, Xiong Xiao, Sining Sun, Wei Rao, Eng~Siong Chng, and Haizhou Li,
\newblock ``Weighted spatial covariance matrix estimation for music based tdoa
  estimation of speech source.,''
\newblock in {\em INTERSPEECH}, 2017, pp. 1894--1898.

\bibitem{180296}
Jie Xiong and Kyle Jamieson,
\newblock ``Arraytrack: A fine-grained indoor location system,''
\newblock in {\em Presented as part of the 10th {USENIX} Symposium on Networked
  Systems Design and Implementation ({NSDI} 13)}, Lombard, IL, 2013, pp.
  71--84, {USENIX}.

\bibitem{bciConvergence}
Mohamed Ibnkahla,
\newblock {\em Adaptive signal processing in wireless communications}, vol.~1,
\newblock CRC Press, 2008.

\bibitem{ReSpeaker}
``Respeaker 6-mic circular array kit for raspberry pi - seeed wiki,'' 2020.

\bibitem{brutti2011inference}
Alessio Brutti, Maurizio Omologo, and Piergiorgio Svaizer,
\newblock ``Inference of acoustic source directivity using environment
  awareness,''
\newblock in {\em Signal Processing Conference, 2011 19th European}. IEEE,
  2011, pp. 151--155.

\bibitem{gollakota2008zigzag}
Shyamnath Gollakota and Dina Katabi,
\newblock ``Zigzag decoding: Combating hidden terminals in wireless networks,''
\newblock in {\em Proceedings of the ACM SIGCOMM 2008 conference on Data
  communication}, 2008, pp. 159--170.

\bibitem{elko1996microphone}
Gary~W Elko,
\newblock ``Microphone array systems for hands-free telecommunication,''
\newblock {\em Speech communication}, vol. 20, no. 3-4, pp. 229--240, 1996.

\bibitem{antonio2011delay}
Sigmund Gudvangen Ragnvald~Otterlei António L. L.~Ramos, Sverre~Holm,
\newblock ``Delay-and-sum beamforming for direction of arrival estimation
  applied to gunshot acoustics,'' 2011.

\bibitem{varma2002time}
Krishnaraj~M Varma,
\newblock {\em Time delay estimate based direction of arrival estimation for
  speech in reverberant environments},
\newblock Ph.D. thesis, Virginia Tech, 2002.

\bibitem{RaspberryPi}
``Raspberry-pi-4-product-brief,'' 2020.

\bibitem{Echo}
``Amazon.com: All-new echo (4th gen),'' 2021.

\bibitem{freire2011doa}
Izabela~L Freire et~al.,
\newblock ``Doa of gunshot signals in a spatial microphone array: performance
  of the interpolated generalized cross-correlation method,''
\newblock in {\em 2011 Argentine School of Micro-Nanoelectronics, Technology
  and Applications}. IEEE, 2011, pp. 1--6.

\bibitem{benesty2004time}
Jacob Benesty, Jingdong Chen, and Yiteng Huang,
\newblock ``Time-delay estimation via linear interpolation and cross
  correlation,''
\newblock {\em IEEE Transactions on speech and audio processing}, vol. 12, no.
  5, pp. 509--519, 2004.

\bibitem{svaizer2012environment}
Piergiorgio Svaizer, Alessio Brutti, and Maurizio Omologo,
\newblock ``Environment aware estimation of the orientation of acoustic sources
  using a line array,''
\newblock in {\em Signal Processing Conference (EUSIPCO), 2012 Proceedings of
  the 20th European}. IEEE, 2012, pp. 1024--1028.

\bibitem{micArrayBCI}
Eric Moulines, Pierre Duhamel, J-F Cardoso, and Sylvie Mayrargue,
\newblock ``Subspace methods for the blind identification of multichannel fir
  filters,''
\newblock {\em IEEE Transactions on signal processing}, vol. 43, no. 2, pp.
  516--525, 1995.

\bibitem{hua1996fast}
Yingbo Hua,
\newblock ``Fast maximum likelihood for blind identification of multiple fir
  channels,''
\newblock {\em IEEE transactions on Signal Processing}, vol. 44, no. 3, pp.
  661--672, 1996.

\bibitem{lin2008blind}
Yuanqing Lin, Jingdong Chen, Youngmoo Kim, and Daniel~D Lee,
\newblock ``Blind channel identification for speech dereverberation using
  l1-norm sparse learning,''
\newblock in {\em Advances in Neural Information Processing Systems (NIPS)},
  2008, pp. 921--928.

\bibitem{lin2007blind}
Yuanqing Lin, Jingdong Chen, Youngmoo Kim, and Daniel~D Lee,
\newblock ``Blind sparse-nonnegative (bsn) channel identification for acoustic
  time-difference-of-arrival estimation,''
\newblock in {\em Applications of Signal Processing to Audio and Acoustics,
  2007 IEEE Workshop on}. IEEE, 2007, pp. 106--109.

\bibitem{nakadai2005sound}
Kazuhiro Nakadai, Hirofumi Nakajima, Kentaro Yamada, Yuji Hasegawa, Takahiro
  Nakamura, and Hiroshi Tsujino,
\newblock ``Sound source tracking with directivity pattern estimation using a
  64 ch microphone array,''
\newblock in {\em Intelligent Robots and Systems, 2005.(IROS 2005). 2005
  IEEE/RSJ International Conference on}. IEEE, 2005, pp. 1690--1696.

\bibitem{nakajima2009real}
Hirofumi Nakajima, Keiko Kikuchi, Toru Daigo, Yutaka Kaneda, Kazuhiro Nakadai,
  and Yuji Hasegawa,
\newblock ``Real-time sound source orientation estimation using a 96 channel
  microphone array,''
\newblock in {\em Intelligent Robots and Systems, 2009. IROS 2009. IEEE/RSJ
  International Conference on}. IEEE, 2009, pp. 676--683.

\bibitem{niwa2010estimation}
Kenta Niwa, Yusuke Hioka, Sumitaka Sakauchi, Ken'ichi Furuya, and Yoichi
  Haneda,
\newblock ``Estimation of sound source orientation using eigenspace of spatial
  correlation matrix,''
\newblock in {\em Acoustics Speech and Signal Processing (ICASSP), 2010 IEEE
  International Conference on}. IEEE, 2010, pp. 129--132.

\bibitem{sit2012direction}
Yoke~Leen Sit, Christian Sturm, Johannes Baier, and Thomas Zwick,
\newblock ``Direction of arrival estimation using the music algorithm for a
  mimo ofdm radar,''
\newblock in {\em 2012 IEEE radar conference}. IEEE, 2012, pp. 0226--0229.

\bibitem{bialer2019performance}
Oded Bialer, Noa Garnett, and Tom Tirer,
\newblock ``Performance advantages of deep neural networks for angle of arrival
  estimation,''
\newblock in {\em ICASSP 2019-2019 IEEE International Conference on Acoustics,
  Speech and Signal Processing (ICASSP)}. IEEE, 2019, pp. 3907--3911.

\bibitem{karanam2018magnitude}
Chitra~R Karanam, Belal Korany, and Yasamin Mostofi,
\newblock ``Magnitude-based angle-of-arrival estimation, localization, and
  target tracking,''
\newblock in {\em 2018 17th ACM/IEEE International Conference on Information
  Processing in Sensor Networks (IPSN)}. IEEE, 2018, pp. 254--265.

\bibitem{haider2018listeer}
Muhammad~Kumail Haider, Yasaman Ghasempour, Dimitrios Koutsonikolas, and
  Edward~W Knightly,
\newblock ``Listeer: Mmwave beam acquisition and steering by tracking indicator
  leds on wireless aps,''
\newblock in {\em Proceedings of the 24th Annual International Conference on
  Mobile Computing and Networking}, 2018, pp. 273--288.

\bibitem{ghasempour2020single}
Yasaman Ghasempour, Rabi Shrestha, Aaron Charous, Edward Knightly, and Daniel~M
  Mittleman,
\newblock ``Single-shot link discovery for terahertz wireless networks,''
\newblock {\em Nature communications}, vol. 11, no. 1, pp. 1--6, 2020.

\bibitem{oumar2012comparison}
Ousmane~Abdoulaye Oumar, Ming~Fei Siyau, and Tariq~P Sattar,
\newblock ``Comparison between music and esprit direction of arrival estimation
  algorithms for wireless communication systems,''
\newblock in {\em The First International Conference on Future Generation
  Communication Technologies}. IEEE, 2012, pp. 99--103.

\bibitem{xie2019md}
Yaxiong Xie, Jie Xiong, Mo~Li, and Kyle Jamieson,
\newblock ``md-track: Leveraging multi-dimensionality for passive indoor wi-fi
  tracking,''
\newblock in {\em The 25th Annual International Conference on Mobile Computing
  and Networking}, 2019, pp. 1--16.

\bibitem{Kotaru:2015:SDL:2785956.2787487}
Manikanta Kotaru, Kiran Joshi, Dinesh Bharadia, and Sachin Katti,
\newblock ``Spotfi: Decimeter level localization using wifi,''
\newblock in {\em Proceedings of the 2015 ACM Conference on Special Interest
  Group on Data Communication (SIGCOMM)}, New York, NY, USA, 2015, pp.
  269--282, ACM.

\bibitem{yun2017strata}
Sangki Yun, Yi-Chao Chen, Huihuang Zheng, Lili Qiu, and Wenguang Mao,
\newblock ``Strata: Fine-grained acoustic-based device-free tracking,''
\newblock in {\em Proceedings of the 15th annual international conference on
  mobile systems, applications, and services}, 2017, pp. 15--28.

\bibitem{sallai2011weapon}
Janos Sallai, Will Hedgecock, Peter Volgyesi, Andras Nadas, Gyorgy Balogh, and
  Akos Ledeczi,
\newblock ``Weapon classification and shooter localization using distributed
  multichannel acoustic sensors,''
\newblock {\em Journal of systems architecture}, vol. 57, no. 10, pp. 869--885,
  2011.

\bibitem{Lazik:2015:DAA}
Patrick Lazik, Niranjini Rajagopal, Oliver Shih, Bruno Sinopoli, and Anthony
  Rowe,
\newblock ``Demo: Alps -- the acoustic location processing system,''
\newblock in {\em Proceedings of the 13th ACM Conference on Embedded Networked
  Sensor Systems (SENSYS)}, 2015.

\bibitem{dokmanic2013acoustic}
Ivan Dokmani{\'c}, Reza Parhizkar, Andreas Walther, Yue~M Lu, and Martin
  Vetterli,
\newblock ``Acoustic echoes reveal room shape,''
\newblock {\em Proceedings of the National Academy of Sciences}, vol. 110, no.
  30, pp. 12186--12191, 2013.

\bibitem{rossi2013roomsense}
Mirco Rossi, Julia Seiter, Oliver Amft, Seraina Buchmeier, and Gerhard
  Tr{\"o}ster,
\newblock ``Roomsense: an indoor positioning system for smartphones using
  active sound probing,''
\newblock in {\em Proceedings of the 4th Augmented Human International
  Conference}, 2013, pp. 89--95.

\end{thebibliography}
\clearpage
\newpage

\end{document}